\documentclass[accepted]{article}

\usepackage{microtype}
\usepackage{graphicx}
\usepackage{subcaption}
\usepackage{booktabs} 

\usepackage{hyperref}

\usepackage{arxiv}

\usepackage{amsmath}
\usepackage{amssymb}
\usepackage{mathtools}
\usepackage{amsthm}

\usepackage[capitalize,noabbrev]{cleveref}

\usepackage{xspace}
\usepackage{bbm}
\usepackage{breqn}
\usepackage{tikz}
\usepackage{adjustbox}
\usepackage{ulem}
\usetikzlibrary{arrows, decorations.pathmorphing, positioning,fit,petri, shapes}

\usepackage{enumitem}
\usepackage{comment}

\theoremstyle{plain}

\theoremstyle{definition}

\theoremstyle{remark}

\icmltitlerunning{Image Compression with Implicit Local Likelihood Models}

\usepackage[textsize=tiny]{todonotes}

\def\mypar#1{\vspace{0mm}{\noindent\bf #1.}\hspace{1mm}}

\newcommand{\Ours}{MS-ILLM\xspace}
\newcommand{\Ourhific}{MS-PatchGAN\xspace}
\newcommand{\hific}{HiFiC\xspace}

\newcommand{\image}{x}
\newcommand{\recon}{\hat{x}}
\newcommand{\latent}{y}
\newcommand{\encoder}{f_{\varphi}}
\newcommand{\decoder}{h_{\upsilon}}
\newcommand{\entropy}{g_{\omega}}
\newcommand{\discr}{D_{\phi}}
\newcommand{\labeler}{u}

\DeclareMathOperator*{\argmin}{arg\,min}

\begin{document}

\twocolumn[
\icmltitle{Improving Statistical Fidelity for Neural Image Compression\\
with Implicit Local Likelihood Models}




\begin{icmlauthorlist}
\icmlauthor{Matthew Muckley}{fair}
\icmlauthor{Alaaeldin El-Nouby}{fair,inria}
\icmlauthor{Karen Ullrich}{fair}
\icmlauthor{Herv\'e J\'egou}{fair}
\icmlauthor{Jakob Verbeek}{fair}
\end{icmlauthorlist}

\icmlaffiliation{inria}{Inria}
\icmlaffiliation{fair}{Meta AI}

\icmlcorrespondingauthor{Matthew Muckley}{mmuckley@meta.com}

\icmlkeywords{Machine Learning, ICML}

\vskip 0.3in
]



\printAffiliationsAndNotice{}  

\begin{abstract}
Lossy image compression aims to represent images in as few bits as possible while maintaining fidelity to the original.
Theoretical results indicate  that optimizing distortion metrics such as PSNR or MS-SSIM  necessarily leads to a discrepancy in the statistics of original images from those of reconstructions, in particular at low bitrates, often manifested by the blurring of the compressed images. 
Previous work has leveraged adversarial discriminators to improve statistical fidelity.
Yet these binary discriminators adopted from generative modeling tasks may not be ideal for image compression. 
In this paper, we introduce a non-binary discriminator that is conditioned on quantized local image representations obtained via  VQ-VAE autoencoders.
Our evaluations on the CLIC2020, DIV2K and Kodak datasets show that our discriminator is more effective for jointly optimizing distortion (e.g., PSNR) and statistical fidelity (e.g., FID) than the PatchGAN of the state-of-the-art HiFiC model.
On CLIC2020, we obtain the same FID as HiFiC with 30-40\% fewer bits.
\end{abstract}

\begin{figure}[ht]
    \begin{center}
    \centering
    \includegraphics[width=0.45\textwidth]{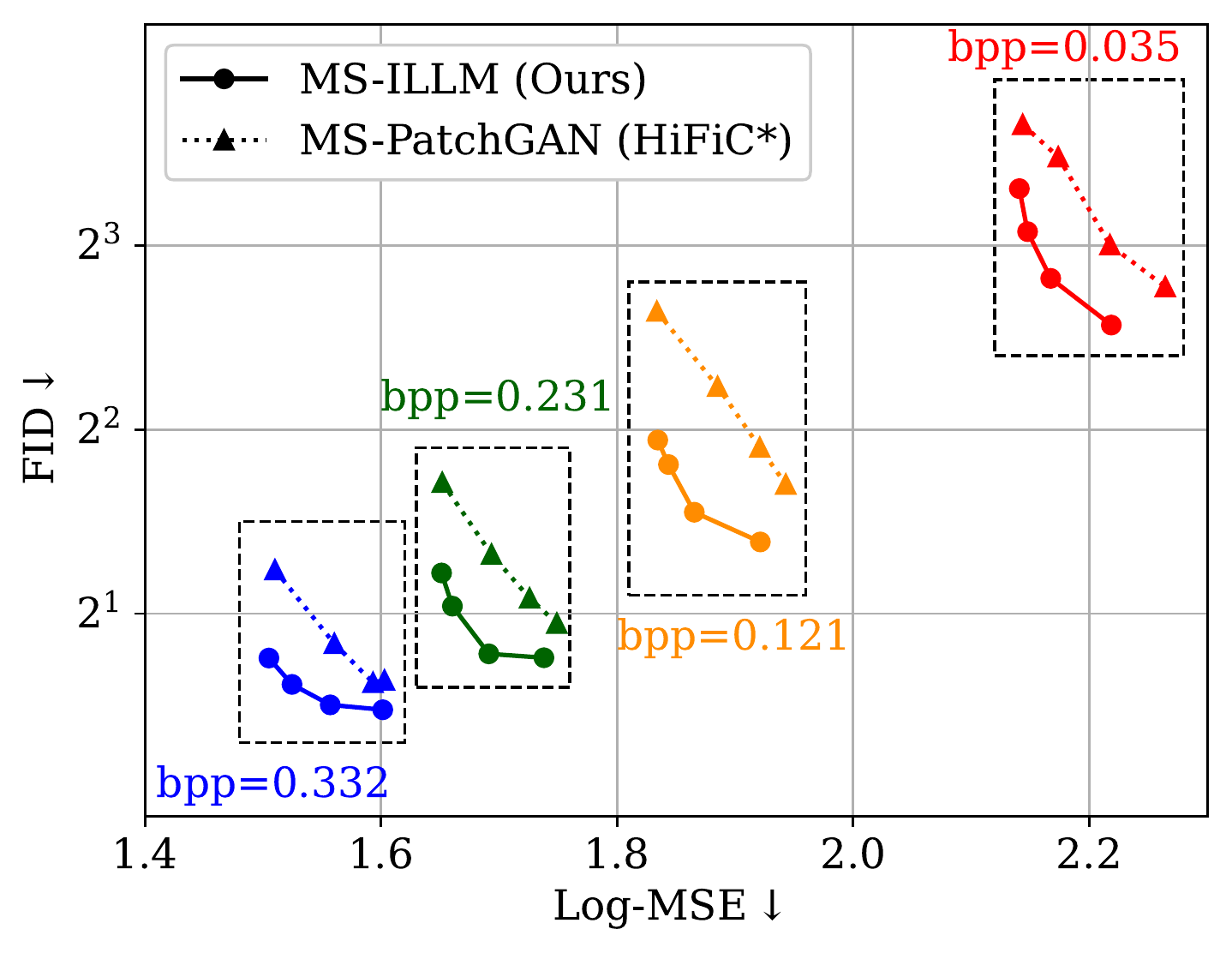}
    \caption{Comparison of distortion vs. statistical fidelity tradeoff across different bitrates. As the bitrate decreases, both distortion (as measured by MSE) and statistical fidelity (as measured by FID) degrade. Throughout all compression regimes, our discriminator achieves better trade-offs between distortion and statistical fidelity than the competing PatchGAN discriminator, as used in HiFiC.
    }
    \label{fig:teaser}
    \end{center}
\end{figure}

\section{Introduction}

The principal task for designing digital image compression systems is to build functions that transform images into the fewest amount of bits while maintaining a fixed, predefined distortion level.
For much of digital compression history, efforts involved designing three components: 
(1) an autoencoding transformation function, such as the discrete cosine transform (DCT)~\citep{ahmed1974discrete}, 
(2) a scheme for lossy quantization of the transform coefficients, and 
(3) a lossless entropy coder for the quantized coefficients.
Arithmetic coding~\cite{rissanen1976ibm,pasco1976source} solves (3) by achieving theoretical bounds~\cite{shannon1948mathematical} for lossless compression.
For the autoencoding transform in (1), handcrafted transforms such as DCT or wavelets~\cite{antonini1992image,le1988sub} can be used.
Combinations of these with carefully-designed  quantization tables for step (2) form the backbone of  compression standards such as JPEG and JPEG2000.

Much of the benefit of JPEG2000 over JPEG was acquired by improving the quality of the autoencoding transform.
In this respect, there is great potential for neural networks to obtain further improvements as end-to-end image compression systems, as has been actively explored, see 
e.g.~\citep{balle2018variational,minnen2018joint,cheng2020joint,mentzer2020high}.
Deep learning methods have now surpassed both old standards (e.g., JPEG and JPEG2000) and new standards (e.g., BPG~\cite{ballard_bpg}) in rate-distortion performance.

A common thread in the design of both traditional image compressors as well as deep learning ones is the rate-distortion optimization objective.
In most cases, methods utilize a handcrafted definition of distortion such as peak-signal-to-noise ratio (PSNR) or multi-scale structural similarity index (MS-SSIM)~\cite{wang2003multiscale,wang2004image}.
However, it can be shown theoretically that especially for very low rates, optimizing for distortion necessarily pulls the statistics of the output samples away from the true distribution~\cite{blau2019rethinking}, typically via blurring or smoothing.
Deep learning compressors that incorporate GANs~\citep{agustsson2019generative,mentzer2020high} mitigate this by balancing distortion loss with an adversarial discriminator that attempts to align the compressed image distribution with the true distribution.
More recent work has shown that the balancing of distortion and statistical fidelity can be controlled at decode time to determine how many details are synthesized~\citep{agustsson2023multi}.
HiFiC~\citep{mentzer2020high} demonstrated that the use of GANs can lead to great benefits from the perspective of human observers, with HiFiC being preferred to BPG even when using half the bits.
Similar results were observed for video compression~\citep{mentzer2022neural}.

While effective, previous approaches for improving statistical fidelity in image compression have primarily relied on discriminator designs from the image generation literature.
Image generation models were designed to model global image distributions.
For compression, the task is different in that the purpose of an adversarial discriminator is a much smaller projection from one image to another, typically from a blurry image to a sharpened image on the statistical manifold of the original natural images.
This process is primarily governed by detail synthesis.
For this reason, we opt to adapt the design of the adversarial training to emphasize this locality.
At a high level, our proposal quantizes all possible images to local neighborhoods, thus aligning the discriminator modeling with its task in compression.

Our contributions are as follows:
\begin{enumerate}
    \item We introduce a new adversarial discriminator based on VQ-VAE autoencoders~\citep{oord2017neural,razavi19nips}. Our new discriminator optimizes likelihood functions in the neighborhood of local images, which we call an ``implicit local likelihood model'' (ILLM). We combine our discriminator with the Mean-Scale Hyperprior~\citep{minnen2018joint} neural compression architecture to create a new compressor that we call a Mean-Scale-ILLM (\Ours).
    \item We perform experiments with \Ours over the CLIC2020, DIV2K, and Kodak  datasets, where we demonstrate that we can surpass the statistical fidelity scores of HiFiC (as measured by FID) without sacrificing PSNR, see Figure~\ref{fig:teaser}.
    \item We ablate our designs over latent dimensions for the VQ-VAE labeler and the U-Net discriminator to validate our architectural design choices.
\end{enumerate}

\section{Related work}
\label{sec:related}

\mypar{Distortion and divergence metrics}
From early on, the standard mean-squared error as a measure of distortion has been criticized for misaligning with human perception of distortion~\citep{snyder1985image}. The (multiscale) structural similarity index measure (MS-SSIM) was proposed to fix this misalignment~\citep{wang2003multiscale,wang2004image} by comparing the statistics of local image patches at multiple resolutions.
Neural alternatives include metrics derived from comparing feature maps of deep networks.
For example, the learned perceptual image patch similarity (LPIPS)~\citep{zhang2018unreasonable} is based on features from a VGG~\citep{simonyan2014very} or AlexNet~\citep{krizhevsky2012imagenet} classifier. The perceptual information metric (PIM) is based on features from unsupervised training~\cite{bhardwaj2020unsupervised}.
Another class of measures includes no-reference measures such as NIQE~\cite{mittal2012making}, FID~\cite{heusel2017gans},  and KID~\cite{binkowski2018demystifying}.
These consider the distributional alignment of the reconstructions and do not measure the distance to the reference data.
More recently, approaches have been developed to learn distortion metrics indirectly via contrastive learning~\cite{dubois2021lossy}. This work relates to other task-centric distortion metrics such as~\cite{tishby2000information,alemi2016deep, torfason2018towards,singh2020end,matsubara2022supervised}.

\mypar{Neural image compression methods}
Optimizing for the standard handcrafted distortion metrics with neural networks can give promising rate-distortion performance~\citep{balle2018variational,minnen2018joint,cheng2020joint,elnouby2023image,he2022elic}.
Optimizing for perceptual metrics is more difficult, as it results in significant compression artifacts~\cite{balle2018variational,ledig2017photo,mentzer2020high,ding2021comparison}.
Thus, in practice often a weighted sum between the loss of a (conditional) GAN and a handcrafted metric such as MSE/MS-SSIM can provide perceptual benefits with stability~\cite{mentzer2020high,agustsson2019generative}. 
One can also use other divergences such as the Wasserstein distance~\cite{tschannen2018deep} or KL-divergence of a deep latent variable model~\citep{theis2022lossy,yang2022lossy,ghouse2023residual}.

\mypar{Rate-distortion-perception tradeoff}
The theoretical limits of the rate-perception-distortion trade-off were investigated by \citet{blau2018perception}. A key finding is that at a given rate, improving the distortion comes at a cost of decreasing the perceptual quality of an image. Later studies have investigated the rate-distortion-perception trade-off, finding that realism generally comes at the expense of rate-distortion~\cite{blau2019rethinking,qian2022rate}, which has also been demonstrated empirically~\cite{theis2022lossy,yang2022lossy}. 
Specifically, perfect realism can be achieved with at most two-fold increase in MSE~\cite{yan2021perceptual,blau2019rethinking}.

\mypar{Minimizing distributional divergence}
One class of divergences that can be used for two-sample hypothesis testing are f-divergences, also known as Ali-Silvey divergences \citep{ali1966general} or Csiszar's $\phi$-divergences \citep{csiszar1967information}. These divergences are connected to the problem of two-sample hypothesis testing because they represent an integrated Bayes risk through their relationship to the density ratio \citep{liese2008f}.
\citet{nowozin2016f} developed a class of generative models, so called f-GANs based on that insight, including the Jensen-Shannon Divergence.
Other GAN architectures have been designed, such as c-GANs~\citep{mirza2014conditional} that  condition image generation on class labels.
Corresponding models exist for when the labels are expanded to spatial semantic maps, such as OASIS~\citep{sushko2022oasis}.
Our work is related to OASIS, the primary differences are 1) they apply their model to the task of semantic image generation, whereas we are doing compression and 2) they use a pixel-space semantic map, whereas we use a latent-space semantic map based on a VQ-VAE~\cite{oord2017neural,razavi19nips}.

\section{Background}
\label{sec:background}
In this section we review rate-distortion theory~\citep{shannon1948mathematical}, as well as more recent work on the rate-distortion-perception tradeoff~\citep{blau2019rethinking}.
At the end of the section, we review how these theories can be applied to the design of neural codecs.

\subsection{Notation}

Throughout this work, we assume $(\Omega, \mathcal{F}, \mathbb{P})$ to be a probability space where $\Omega$ is the sample space, $\mathcal{F}$ is the event space, and $\mathbb{P}$ denotes the probability function such that $X: \Omega \to \mathcal{X}$ is a random variable (r.v.) defined on the space.
Equivalently, $Y: \Omega \to \mathcal{Y}$. 
We will use capital letters for random variables, e.g.\ $X$; lower case letters for their realizations, e.g.\ $x\in \mathcal{X}$; $P_X$ is a distribution of $X$; and $p_X$ is the probability mass function of $P_X$.  
We will denote conditional distributions as $P_{X|Y}$, which we think of as a collection of probability measures on $\mathcal{X}$, for each value $y$ there exists $P_{X|Y=y}$. 
Expectations will be denoted as $\mathbb{E}_{x\sim P_X} \left[q(x)\right]$, or abbreviated as $\mathbb{E}\left[q(x)\right]$.

\subsection{Rate-distortion-perception theory}

The goal of lossy compression is to store the outcomes $\image \sim P_X$ of a discrete random variable $X$, e.g.\ natural images with as few bits (bit-rate) as possible while simultaneously ensuring that a reconstruction $\recon \sim P_{\hat{X}|x}$ is of a certain quality level no lesser than $\tau$. 
The problem has been formulated more precisely as rate-distortion theory \cite{shannon1948mathematical}. Shannon concludes that the best bit-rate $R$ is characterized by the rate-distortion function;
\begin{align}\small\label{eq:RD-fun}
    &R(\tau) = \min\limits_{P_{\hat{X}|X}} I(\hat{X};X) \\\notag
    &\text{ s.t.  }~ \mathbb{E}_{\image,\recon \sim P_X P_{\hat{X}|x}}\left [ \rho(\recon, \image) \right ] \leq \tau,
\end{align}
where $I(\cdot;\cdot)$ denotes the mutual information, and $\rho(\cdot, \cdot)$ is a 
distortion measure. 
\citet{blau2019rethinking} recently extended the aforementioned rate-distortion function to include an additional constraint that characterizes how well the statistics of the reconstructions resemble statistics of the real data distribution;
\begin{align} \label{eq:RDP-fun}
  d(P_{\hat{X}},P_X) \leq \sigma,
\end{align}
where $d(\cdot, \cdot)$ is some divergence between distributions, e.g.\ the Kulback-Leibler divergence (KLD).
While the authors refer to this as ``perception'', for sake of clarity we refer to this metric as ``statistical fidelity'' to differentiate it from human perception, as well as metrics such as LPIPS~\cite{zhang2018unreasonable} that are considered ``perceptual'' in the literature.
A key finding of \citet{blau2018perception} is that in many cases statistical fidelity comes at the expense of distortion, at constant rate. 

Below, we show how to build a differentiable training objective by approximating and relaxing the constrained rate-distortion-perception function  $R(\tau,\sigma)$. 

\subsection{Lossy codec optimization}

For the purpose of this paper, we will constrain ourselves to lossy compression algorithms (codecs) of the following kind: We assume source symbols $\image$ will be encoded into a (quantized) latent representation $\latent=f(\image)$, $f: \mathcal{X} \rightarrow \mathcal{Y}$. 
Subsequently, we employ an entropy coder\footnote{An entropy coder is a map that solves the lossless compression problem optimally. Please see \citet{cover,mackay} for more details.} $g_{\omega}(\latent)$ with parameters $\omega$ to losslessly compress $\latent$ into its shortest possible bit string.
We write $r(\image):=|g_{\omega}(\latent)|$ to denote the bit rate or length of the binary string generated by $g_{\omega}$.
Since sender and receiver have common knowledge of the entropy coder, the receiver can apply a decoder $h: \mathcal{Y} \rightarrow \mathcal{X}$ to recover the source signal $\recon=h(\latent)=f \circ h(\image)$.
We will refer to the tuple of encoder, decoder and entropy coder as a lossy codec $(f,g_{\omega},h)$.
See Figure~\ref{fig:infoflow} for an overall system diagram.

The goal of optimization is to learn a parameterized lossy codec, $(\encoder,\entropy,\decoder)$, where $\varphi$, $\omega$, and $\upsilon$ denote the parameters of each component.
In our case, the encoder $\encoder$ and  the decoder $\decoder$ are neural networks  and the entropy coder $\entropy$ will be defined by a parameterized approximation of the marginal over representations, in other words we need to learn $P_{Y|\omega}$.
See \citet{balle2017end} for details of the overall structure; we describe ours in Section \ref{sec:method}.

Using Lagrange multipliers to relax \eqref{eq:RD-fun} and \eqref{eq:RDP-fun}, the training objective comes out to be 
\begin{align}\label{eq:objective}
    \mathcal{L}(\varphi,\omega,\upsilon) = 
    &\lambda_r \text{ } \mathbb{E}_{\image \sim P_X}\left [ r _{\varphi,\omega} \left ( \image \right )  \right ]  \\\notag
    + &\lambda_{\rho} \text{ } \mathbb{E}_{\image \sim P_X}\left [ \rho\left( \encoder \circ \decoder(\image),\image \right )  \right ]\\\notag
    + &\lambda_d \text{ } d\left (P_{\hat{X}}, P_X \right ). 
\end{align}
Before we give detailed descriptions of the functional forms of our lossy codec, we need to specify how we can approximate the distributional divergence in practice.

\subsection{Approximating the distributional divergence}

Mechanistically, it is typically straightforward to specify distortion functions for $\rho\left( \encoder \circ \decoder (\image), \image \right )$, but minimizing the divergence term, $d\left (P_{\hat{X}},P_X\right )$ can be move involved.
A standard technique is to use GANs to optimize the symmetric Jensen-Shannon divergence (JSD)~\citep{goodfellow2014generative,nowozin2016f}.
The JSD is a proper divergence measure between distributions, meaning that if there are enough training samples and the model class $P_{\hat{X}}$ is sufficiently rich, $P_X$ can be accurately approximated. 
\citet{goodfellow2014generative} show the JSD is minimized by the well-known GAN minimax optimization problem
\begin{align}\small \label{eq:basicdiscr}
    \min\limits_{\phi} \max\limits_{\varphi,\omega,\upsilon} \text{ } &\mathbb{E}_{x\sim P_{X}} \left [-\text{ log }\discr(\image) \right ] \\ \notag
    &+ \mathbb{E}_{\hat{x}\sim P_{\hat{X}}} \left [-\text{ log} \left (1-\discr(\recon) \right ) \right ],
\end{align}
where $D_{\phi}$ is a parameterized discriminator function (with parameters $\phi$) that estimates if a sample was drawn from the real data distribution and we have used the shorthand $\recon=\encoder \circ \decoder (\image)$.
In neural compression, we assume $P_{\hat{X}}$ to be the marginal over the joint $P_XP_{\hat{X}|x}$.
For computing the empirical risk in (\ref{eq:basicdiscr}), we draw different samples for $x$ and $\hat{x}$.
The sign-flipped generator-loss function from (\ref{eq:basicdiscr}) is
\begin{equation}
    \mathcal{L}_G(\varphi,\omega,\upsilon) = \mathbb{E}_{\hat{x}\sim P_{\hat{X}}} \left [\text{ log} \left (1-\discr(\recon) \right ) \right ],
\end{equation}
which we can use as a drop-in replacement for $d(P_{\hat{X}},P_X)$ in \eqref{eq:objective} (alternating minimization for the discriminator).
This approach has been applied to several neural compression systems~\cite{agustsson2019generative,mentzer2020high}. 
Compared to GANs for image generation, the  task for $P_{\hat{X}}$ is greatly simplified by the  properties of the neural compression task.
For this reason, we redesign the discriminator to reflect the locality of projection needed for compression.

\tikzset
{
  myTrapezium/.pic =
  {
    \draw (0,0) -- (0,\b) -- (\a,\c) -- (\a,-\c) -- (0,-\b) -- cycle ;
    \coordinate (-center) at (\a/2,0);
    \coordinate (-out) at (\a,0);
  },
  myArrows/.style=
  {
    line width=2mm, 
    red,
    -{Triangle[length=1.5mm,width=5mm]},
    shorten >=2pt, 
    shorten <=2pt, 
  }
}
    \def\a{1}  
    \def\b{.5} 
    \def\c{1}  

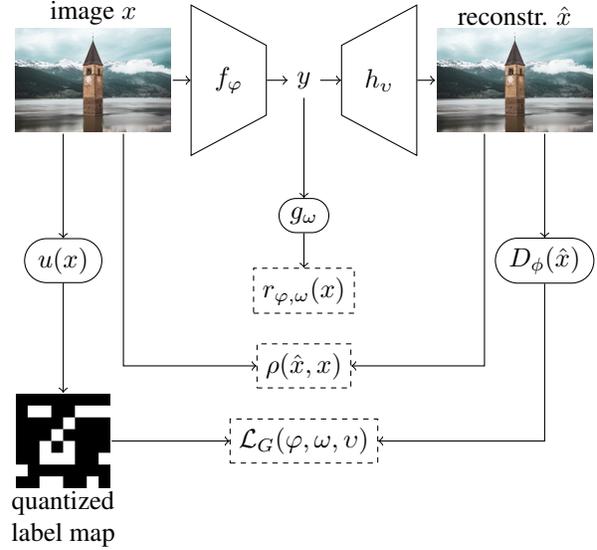
\begin{figure}[ht]
    \centering
    \begin{tikzpicture}
        \node[anchor=center,inner sep=0.5pt] (input) at (0,0) {\includegraphics[width=.25\columnwidth]{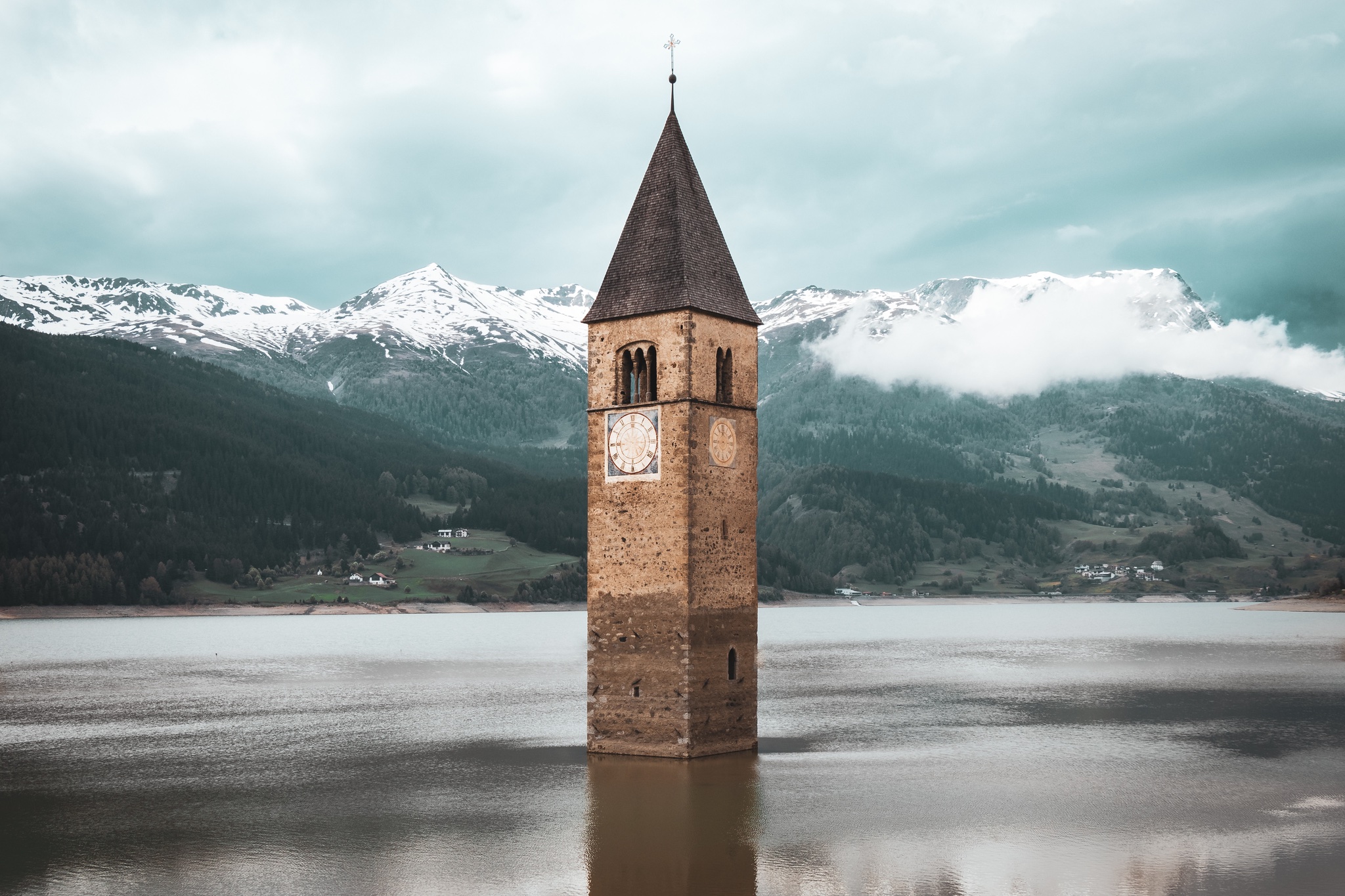}};
        \node[anchor=south,inner sep=0.5pt] (input_label) at (input.north) {image $\image$};

        \node[trapezium,
        draw,
        right of=input,
        node distance=1.8cm,
        trapezium stretches=true,
        minimum height=1cm,
        minimum width=2cm,
        trapezium left angle=120,
        trapezium right angle=120,
        rotate=90] (encoder) {};
        \node at (encoder.center) {{$\encoder$}};

        \node[
        right of=encoder,
        node distance=1cm] (latent) {$\latent$};

        \node[trapezium,
        draw,
        right of=latent,
        node distance=1cm,
        trapezium stretches=true,
        minimum height=1cm,
        minimum width=2cm,
        trapezium left angle=120,
        trapezium right angle=120,
        rotate=-90] (decoder) {};
        \node at (decoder.center) {$\decoder$};

        \node[inner sep=0.5pt,
        right of=decoder,
        node distance=1.8cm] (output) {\includegraphics[width=.25\columnwidth]{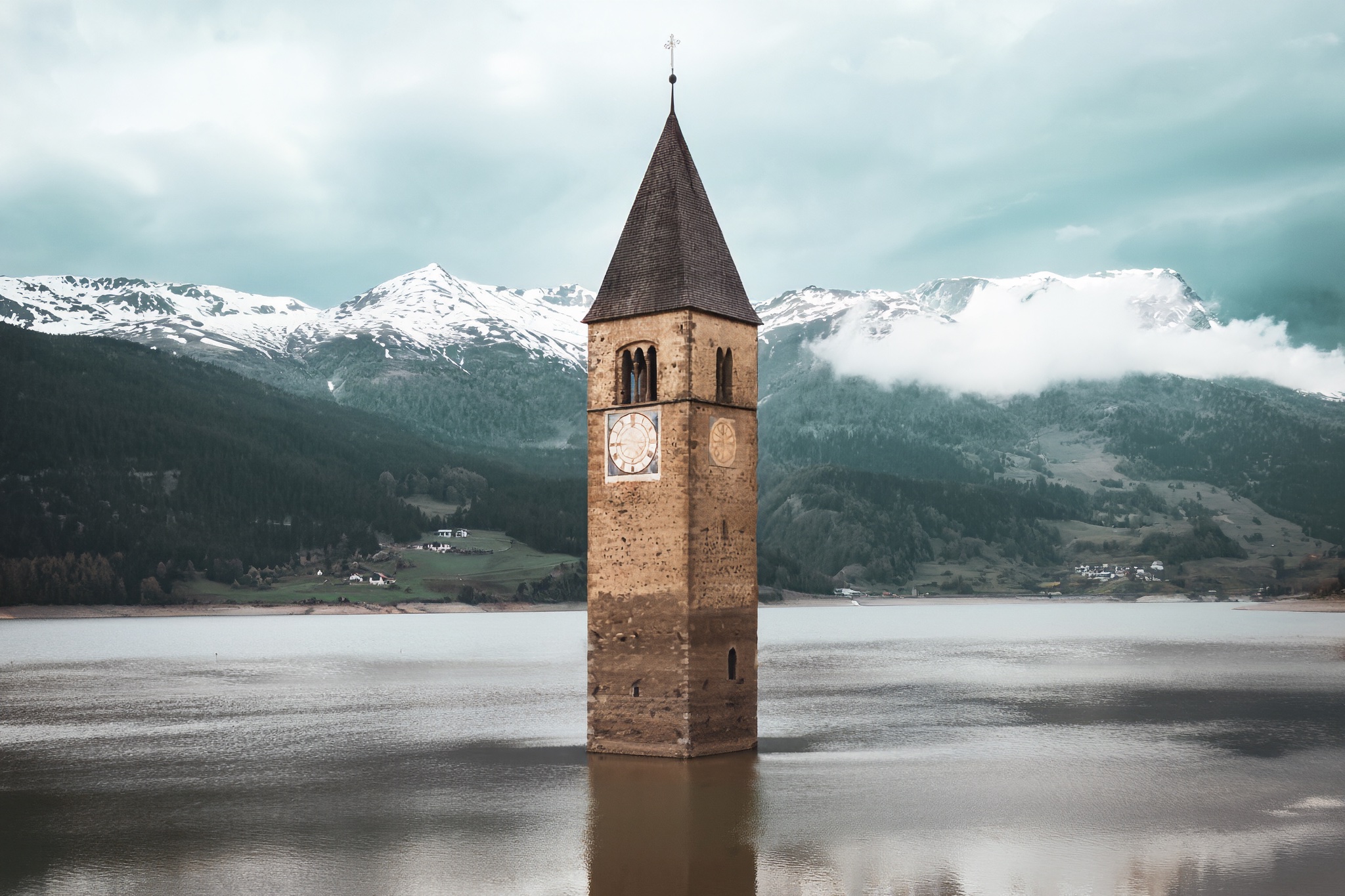}};
        \node[anchor=south,inner sep=0.5pt] (output_label) at (output.north) {reconstr. $\recon$};


        \node[rounded rectangle,
        draw,
        below of=latent,
        node distance=1.8cm] (entropy) {$\entropy$};


        \node[
        draw,
        rectangle,
        dash pattern=on 2pt off 2pt,
        below of=entropy,
        node distance=1cm
        ] (rate) {$r _{\varphi,\omega}(\image)$};

        \node[
        draw,
        rectangle,
        dash pattern=on 2pt off 2pt,
        below of=rate,
        node distance=1cm
        ] (distortion) {$\rho(\recon, \image)$};

        \node[
        draw,
        rectangle,
        dash pattern=on 2pt off 2pt,
        below of=distortion,
        node distance=1cm
        ] (divergence) {$\mathcal{L}_G(\varphi,\omega,\upsilon)$};


        \node[rounded rectangle,
        draw,
        xshift=-0.4cm,
        yshift=-2.4cm,
        node distance=1.8cm] at (input.center) (labeler) {$\labeler(\image)$};

        \node[rounded rectangle,
        draw,
        xshift=0.4cm,
        yshift=-2.4cm,
        node distance=1.8cm] (discriminator) at (output.center) {$\discr(\recon)$};

        \node[inner sep=0.5pt] at (labeler |- divergence) (map) {\includegraphics[width=.15\columnwidth]{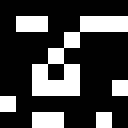}};
        \node[anchor=north,
        inner sep=0.5pt,
        align=center] (map_label) at (map.south) {quantized\\ label map};
        
        \draw[->] (input.east) -> (encoder.north);
        \draw[->] (encoder.south) -> (latent.west);
        \draw[->] (latent.east) -> (decoder.south);
        \draw[->] (decoder.north) -> (output.west);

        \draw[->] (latent.south) -> (entropy.north);
        \draw[->] (entropy.south) -> (rate.north);
        
        \draw[->] (input.south -| labeler) -> (labeler.north);
        \draw[->] (labeler.south) -> (map.north);
        \draw[->] (map.east) -> (divergence.west);

        \draw [->] ([xshift=0.4cm]input.south) -- ([xshift=0.4cm]distortion -| input.south) -> (distortion.west);

        \draw [->] ([xshift=-0.4cm]output.south) -- ([xshift=-0.4cm]distortion -| output.south) -> (distortion.east);

        \draw [->] (output.south -| discriminator) -> (discriminator.north);
        \draw [->] (discriminator.south) -- (divergence.east -| discriminator) -> (divergence.east);
        
    \end{tikzpicture}
    \caption{Overview of our learned lossy compression system with discriminator-based divergence minimization. An image $x$ is encoded and quantized to latent $y$. A likelihood model $g_\omega$ enables entropy coding of $y$ and an estimate of the rate $r _{\varphi,\omega}$. A decoder $h_\upsilon$ converts quantized $y$ back into a compressed image $\hat{x}$. To improve  statistical fidelity, we also train with a discriminator, $\discr$, that attempts to match labels from a pretrained labeler $\labeler$.
    }
    \label{fig:infoflow}
\end{figure}

\section{Method}
\label{sec:method}
Our approach follows that of HiFiC~\citep{mentzer2020high}, with the primary distinction being in a novel proposal for modeling local likelihoods in the neighborhood of the compressed image.
We describe our vector-valued OASIS-type discriminator $\discr$~\citep{sushko2022oasis}, the most crucial adaptation as compared to other neural compression schemes, and point to relevant training specifics.

\subsection{Autoencoder architecture}
Here, we  describe the encoder $\encoder$, decoder $\decoder$ and the latent marginal $P_{Y|\omega}$ as used by the entropy coder, $g_{\omega}$.
Due to the relationship between the rate-distortion function and variational inference, a neural codec can be viewed as a type of variational autoencoder~\cite{balle2017end}.  
In line with previous work in neural compression, we model the data distribution using a two-level hierarchical autoencoder, also known as a hyperprior model~\cite{balle2018variational}.
This model class is named after the governing prior $P_{Y|\omega}$ that itself is modeled as a latent variable model $P_{Y|\omega}=\sum P_{Y|z;\omega}P_{z}$.

Previous work has used different choices for the prior distribution, such as a conditional Gaussian~\cite{balle2018variational,minnen2018joint,mentzer2020high}. 
We adopt the approach of \citet{minnen2018joint} that conditions the means and scales of the Gaussian, and refer  to it as the mean-scale hyperprior model.
For the architecture of the encoder and decoder we follow \citet{mentzer2020high}, who used larger and deeper models than \citet{minnen2018joint}.

\subsection{Implicit local likelihood models}

To develop our likelihood models, we assume that we have access to a labeling vector function, $\labeler: \mathcal{X} \rightarrow \{0, 1\}^{(C+1)\times W \times H }$.
For any $\image$ in our dataset, $\labeler$ outputs a 3D spatially-distributed one-hot target vector map with dimensions $(C+1) \times W \times H$, where $C$ is the number of labels, $W$ is the latent space width, and $H$ is the latent space height.
We reserve the zero-th label as a ``fake'' class to designate  reconstructed images, and use the remaining $C$ classes to label original images.
We define $b_0$ as a one-hot target $(C+1) \times W \times H$ tensor  where  values are $1$ for the zero-th element in the $C$ dimension, effectively the ``fake'' class in standard GAN terminology.

Following \citet{sushko2022oasis}, we can now define the two c-GAN-style adversarial loss functions:
\begin{align}
    \label{eq:oasisdisc} \mathcal{L}_D(\phi) =& \text{ }\mathbb{E}_{\image\sim P_{X}} \left [-\langle \labeler(\image), \log\discr(x) \rangle \right ] \\ &+ \mathbb{E}_{\recon\sim P_{\hat{X}}} \left [-\langle b_0, \log \discr(\recon)  \rangle \right ], \notag \\
    \label{eq:oasisgen} \mathcal{L}_G(\varphi,\omega,\upsilon) =& \text{ }\mathbb{E}_{\recon\sim P_{\hat{X}}} \left [-\langle \labeler(\image), \log  \discr(\recon)  \rangle \right ],
\end{align}
where $\discr(\image)$ is now a vector-valued function, and $\langle \cdot, \cdot \rangle$ denotes the inner product.
Note that rather than just distinguishing original from reconstructed images, here the goal of the discriminator is to distinguish among the $C$ image labels of original images, as well as detecting reconstructed images.
The generator loss corresponds to the non-saturating GAN loss proposed by \citet{goodfellow2014generative}, extended to the multi-label case. 
It  aims, for reconstructed images,  to maximize the discriminator likelihood of the labels of the corresponding real image.

Effectively, this allows the discriminator to use more local information in the label.
We note that HiFiC also includes an alternative form of locality in that it uses the latent as an extra input to the PatchGAN discriminator~\citep{mentzer2020high}.
However, in this case the discriminator still has the opportunity to ignore local information, because locality is not enforced in the loss function.
Conversely, our approach enforces locality by including a latent code in the loss function and inputing this information via backpropagation.
The distinction between forward-conditioning and loss-based conditioning is discussed in the OASIS paper~\citep{sushko2022oasis}, and we refer readers there for further details.

\subsection{Choice of labeling function}

The choice of labeling function influences the success of the method.
Equations~(\ref{eq:oasisdisc}) and (\ref{eq:oasisgen}) allow broad classes of labels, including global image labels when setting $W=H=1$ and spatially-distributed  labels for $H,W>1$.
Since our goal is to enforce locality in the implicit likelihood model, we opt to apply VQ-VAEs~\citep{oord2017neural,razavi19nips}.
Using vector quantization, we partition the latent space of a VQ-VAE autoencoder into $C$ clusters with cluster means $\{m_c\}_{c=1}^C$. 
For original images, we set
 \begin{equation}
     \labeler^{(c,i,j)}(\image) = \begin{cases} 1\quad \text{if }\;c = \argmin\limits_{q=1,\dots,C} \left \| e^{(i,j)} - m_q \right \|_2^2, \\
     0 \quad \text{otherwise},
     \end{cases}
 \end{equation}
where $e^{(i,j)}$ is the channel vector at location $(i,j)$ from the VQ encoder.
The VQ-VAE encoder  partitions the  space of the latent vectors into $C$ Voronoi cells.
Since Voronoi cells are convex, any two points are path connected. 
Further, since our decoder architecture is a continuous mapping, the path connectedness remains even in reconstruction space $\mathcal{X}$.
In other words, with the VQ-VAE approach, locality in the label space implies locality in image space.
This contrasts with other labeling approaches, such as with ImageNet classes.
Without a model for generating latent codes, our approach does not consider unconditional generation, although this could be achieved using autoregressive techniques~\citep{oord2017neural,razavi19nips}.

Our VQ-VAE architecture is based on the VQ-GAN variant~\citep{esser2021taming} with a couple of modifications:
\begin{itemize}[itemsep=1pt,topsep=2pt]
    \item We use ChannelNorm~\citep{mentzer2020high} instead of GroupNorm to improve normalization statistical stability across different image regions.
    \item We use XCiT~\citep{elnouby2021xcit} for the attention layers to improve compute efficiency.
\end{itemize}
Given this architecture, the number of likelihood functions is governed by the spatial size of the latent space ($W\times H)$ and codebook size $C$ of the VQ-VAE.
A larger latent space and codebook will lead to smaller likelihood neighborhoods.
Unless specified otherwise,  we utilize a $32 \times 32$ latent space for images of size $256 \times 256$ and a codebook size of 1024.
We did not observe a huge variation in results depending on these parameters (see ablations in Section~\ref{sec:ablations}).

\subsection{Discriminator architecture}

For the discriminator we use the U-Net architecture~\cite{ronneberger2015u} previously proposed  by \citet{sushko2022oasis} in the context of  semantic image synthesis. 
In their case, the discriminator aims to label pixels of real images with the corresponding label in the conditioning semantic segmentation map, while labeling pixels in generated images as ``fake''. 
In our case, rather than predicting manually annotated semantic classes, the discriminator predicts among the labels provided by the labeling function $u$. 

Our U-Net variant uses LeakyReLU~\citep{xu2015empirical} for the activations and is built on top of residual blocks rather than the feed-forward blocks of the original~\citep{ronneberger2015u}.
Since our latent resolution is different from the image resolution, we cut the output path of the U-Net at the level of the $32 \times 32$ latent provided by $\labeler(\image)$.
Contrary to the OASIS discriminator architecture, we do not use  normalization for the convolutional layers.
We considered several types of normalization, including spectral~\citep{miyato2018spectral} and instance normalization~\citep{ulyanov2016instance}, but we found that no normalization at all was most effective.
We show ablations for the normalization layers in Section~\ref{sec:ablations}.

\subsection{Training}
We utilize the two-stage training process of HiFiC~\cite{mentzer2020high}.
In the first stage, we train the autoencoder (i.e., encoder $\encoder$, entropy coder $\entropy$, and decoder $\decoder$) without the discriminator for 1M steps.
In pretraining the autoencoder we observed two phenomena.
First, there were very large gradients at the start of training, leading to high variance.
Second, in late training, the model requires very small gradient steps in order to optimize for the last few dB of PSNR.
For these reasons, we depart from the learning rate schedule of HiFiC: we begin with linear warmup~\citep{liu2020variance} for 10,000 steps and adopt a cosine learning rate decay for the rest of training.
We set a peak learning rate of $3 \times 10^{-4}$ and train using the AdamW optimizer~\citep{loshchilov2017decoupled} with a weight decay of $5 \times 10^{-5}$.
We adopt the rate targeting strategy of HiFiC for six bitrates.

In addition, we pass the quantized latents to the decoder with backpropagation via the straight-through estimator~\cite{theis2017lossy} to ensure the decoder sees the same values at training and test-time.
The models from this first training stage are used in all subsequent fine-tuning experiments.
Our procedure for training the quantizing model $\labeler(\image)$ is almost identical, except for setting a smaller value of the MSE parameter in $\rho$ (see Appendix), aligning features more with LPIPS~\citep{zhang2018unreasonable}.

In the second stage we finetune the decoder, $h_\upsilon$, part of the autoencoder using the full rate-distortion-perception loss of (\ref{eq:RDP-fun}).
We use a learning rate of $4 \times 10^{-4}$ for the U-Net discriminator and $1 \times 10^{-4}$ for the generator, which are the same values used in OASIS~\citep{sushko2022oasis}.
We tried tuning the discriminator and generator learning rates around these points, but we found that for higher bitrates in particular it was necessary for the discriminator to train faster than the generator, and the ($4\times10^{-4}$, $1\times10^{-4}$) tuple represented a sweet spot for achieving this.
For fine-tuning, we used the AdamW~\citep{loshchilov2017decoupled} optimizer with the same parametes of pretraining, except we lower the betas to (0.5, 0.9).

We refer to the overall model from this training process ``\Ours'', for Mean \& Scale Hyperprior fine-tuned with the Implicit Local Likelihood Model, to emphasize the locality of the discriminator vs.\ previous methods.

\section{Experiments}
\label{sec:experiments}
\begin{figure*}[ht]
    \centering
    \begin{subfigure}{.8\textwidth}
        \includegraphics[width=\columnwidth]{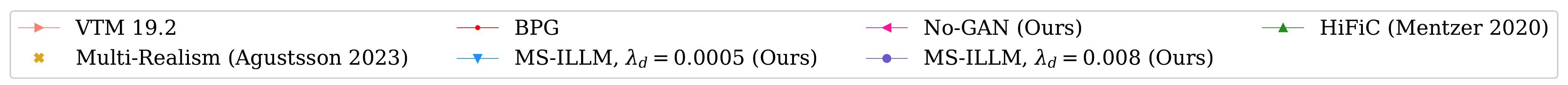}
    \end{subfigure}
    
    \begin{subfigure}{.33\textwidth}
        \includegraphics[width=\columnwidth]{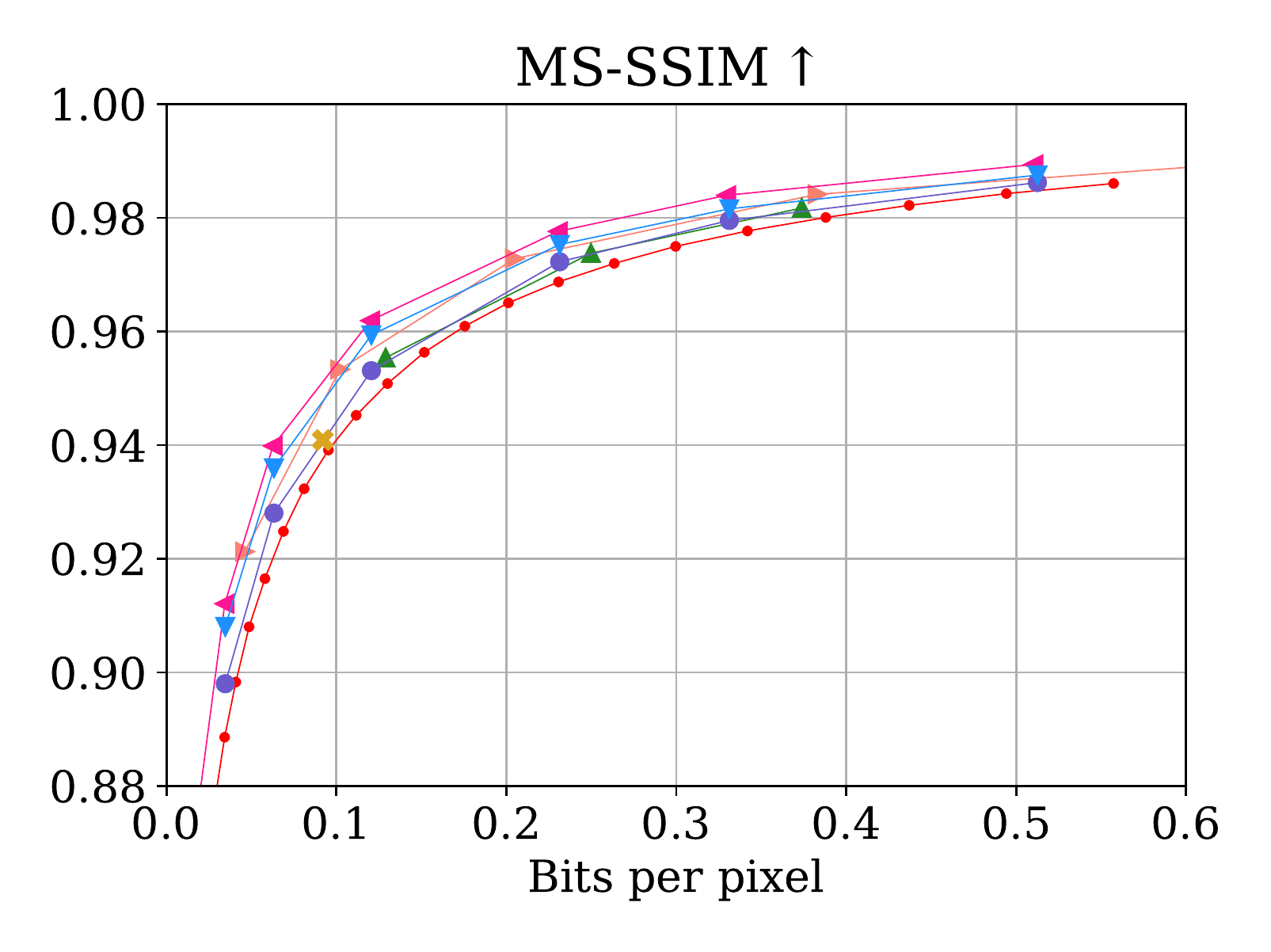}
    \end{subfigure}
    \hfill
    \begin{subfigure}{.33\textwidth}
        \includegraphics[width=\columnwidth]{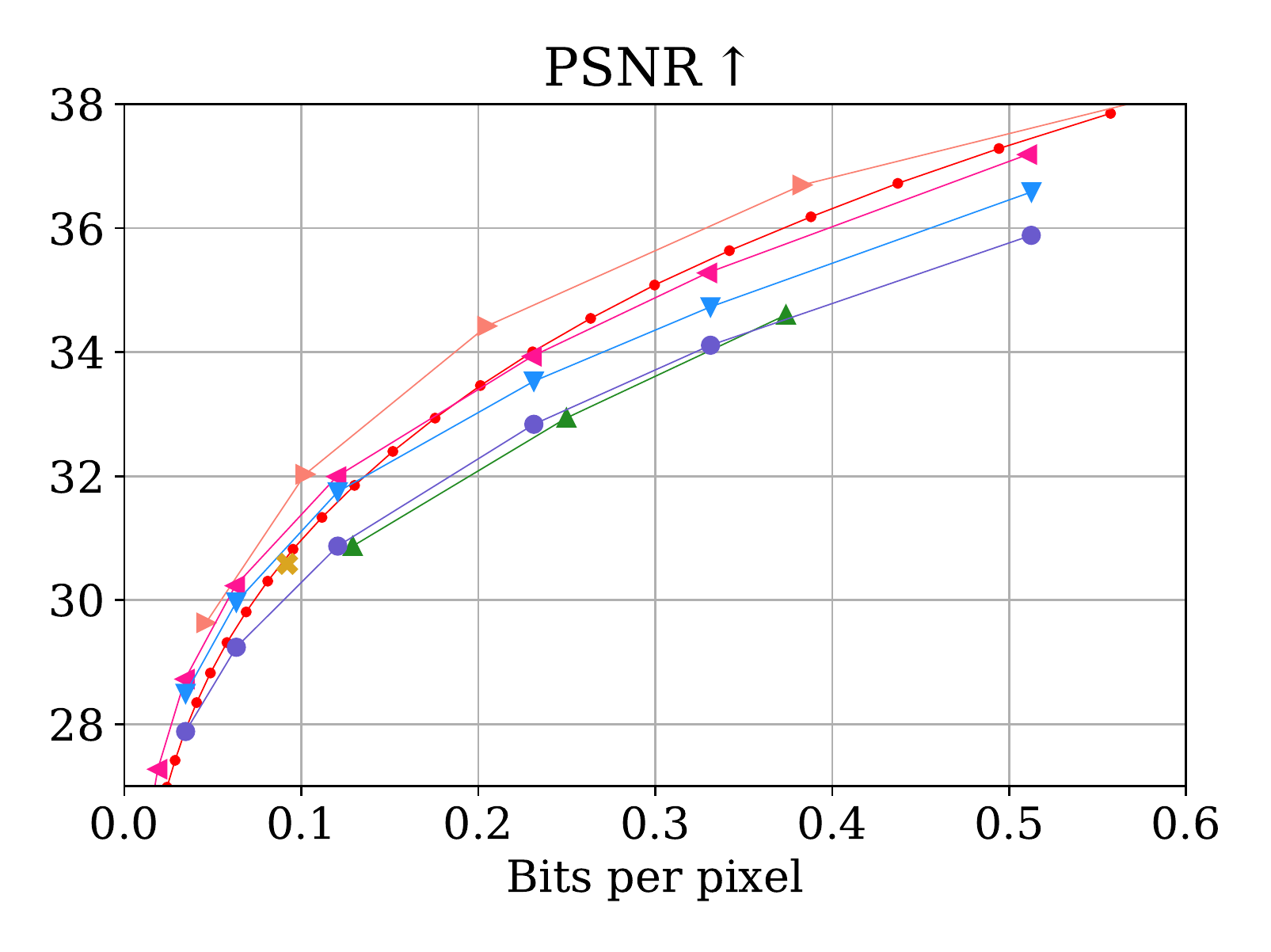}
    \end{subfigure}
    \hfill
    \begin{subfigure}{.33\textwidth}
        \includegraphics[width=\columnwidth]{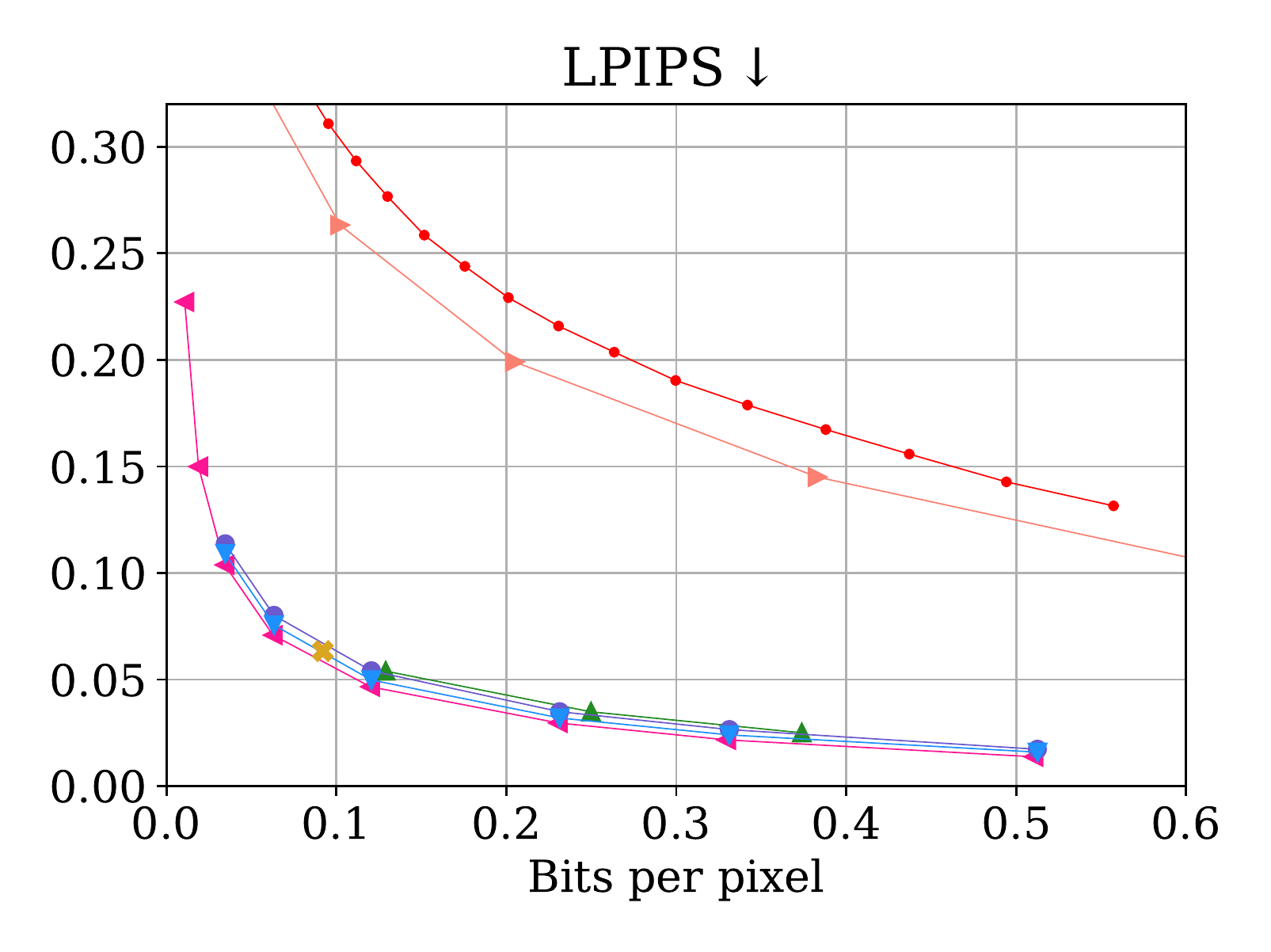}
    \end{subfigure}

    \begin{subfigure}{.33\textwidth}
        \includegraphics[width=\columnwidth]{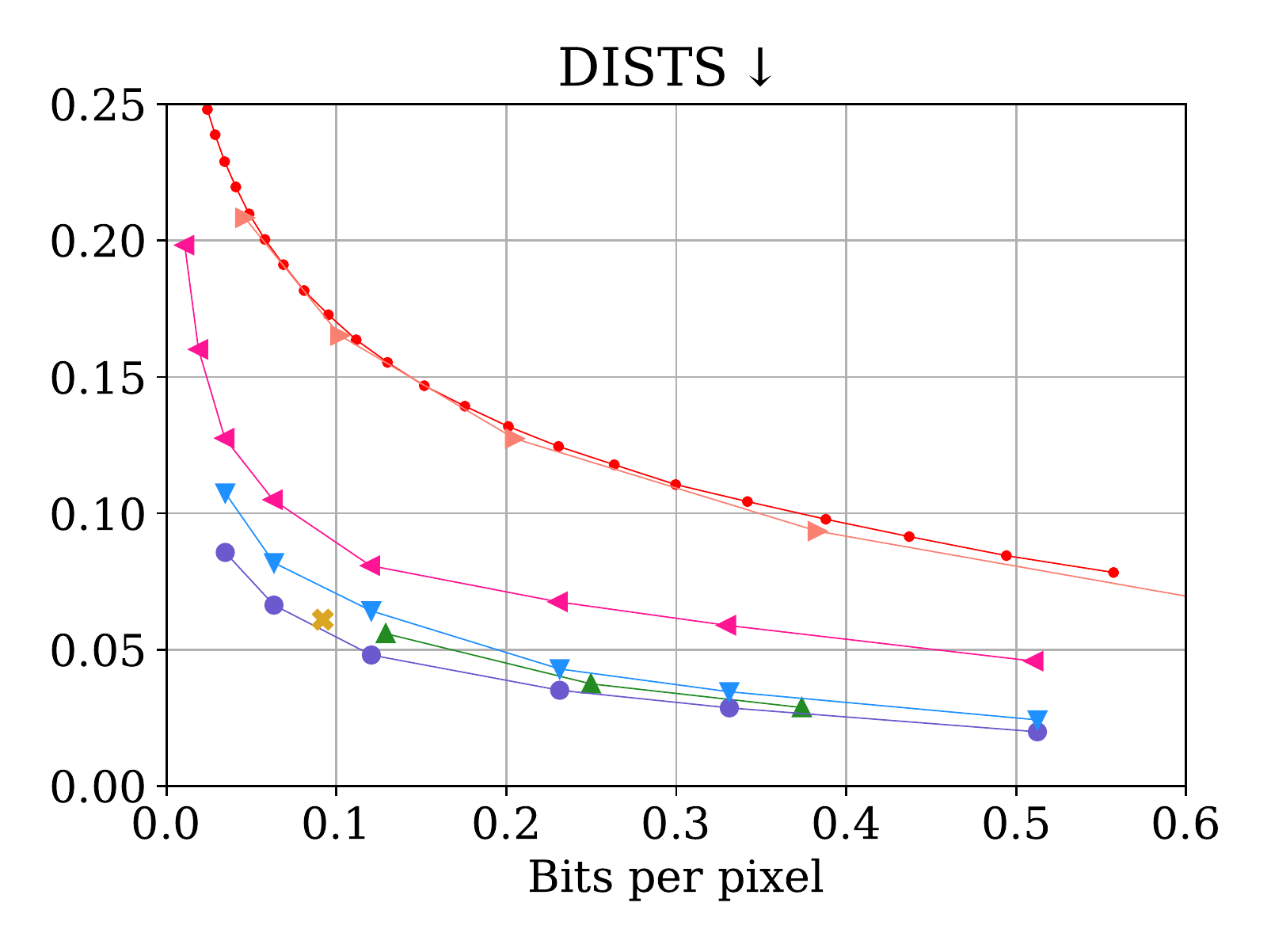}
    \end{subfigure}
    \hfill
    \begin{subfigure}{.33\textwidth}
        \includegraphics[width=\columnwidth]{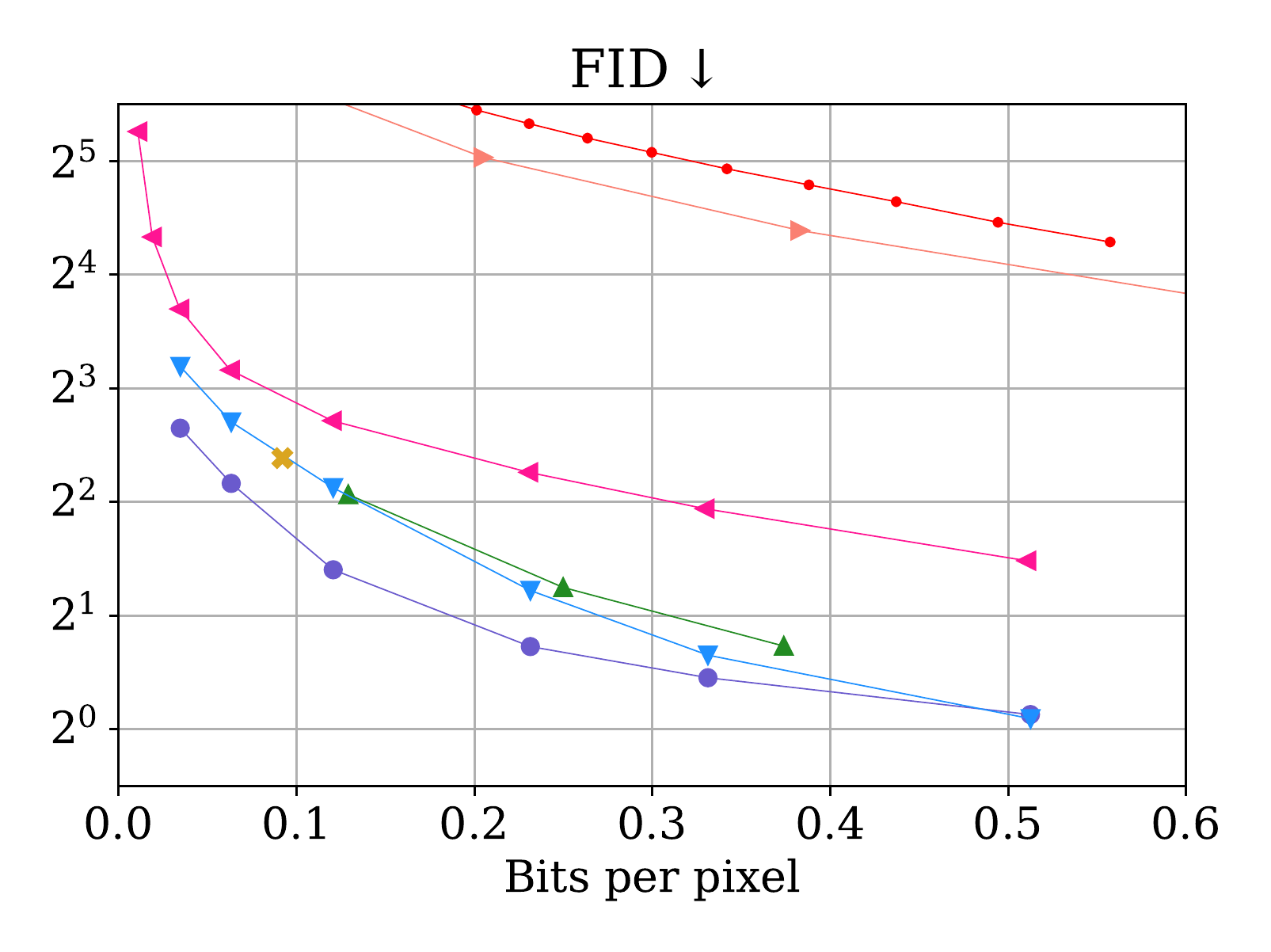}
    \end{subfigure}
    \hfill
    \begin{subfigure}{.33\textwidth}
        \includegraphics[width=\columnwidth]{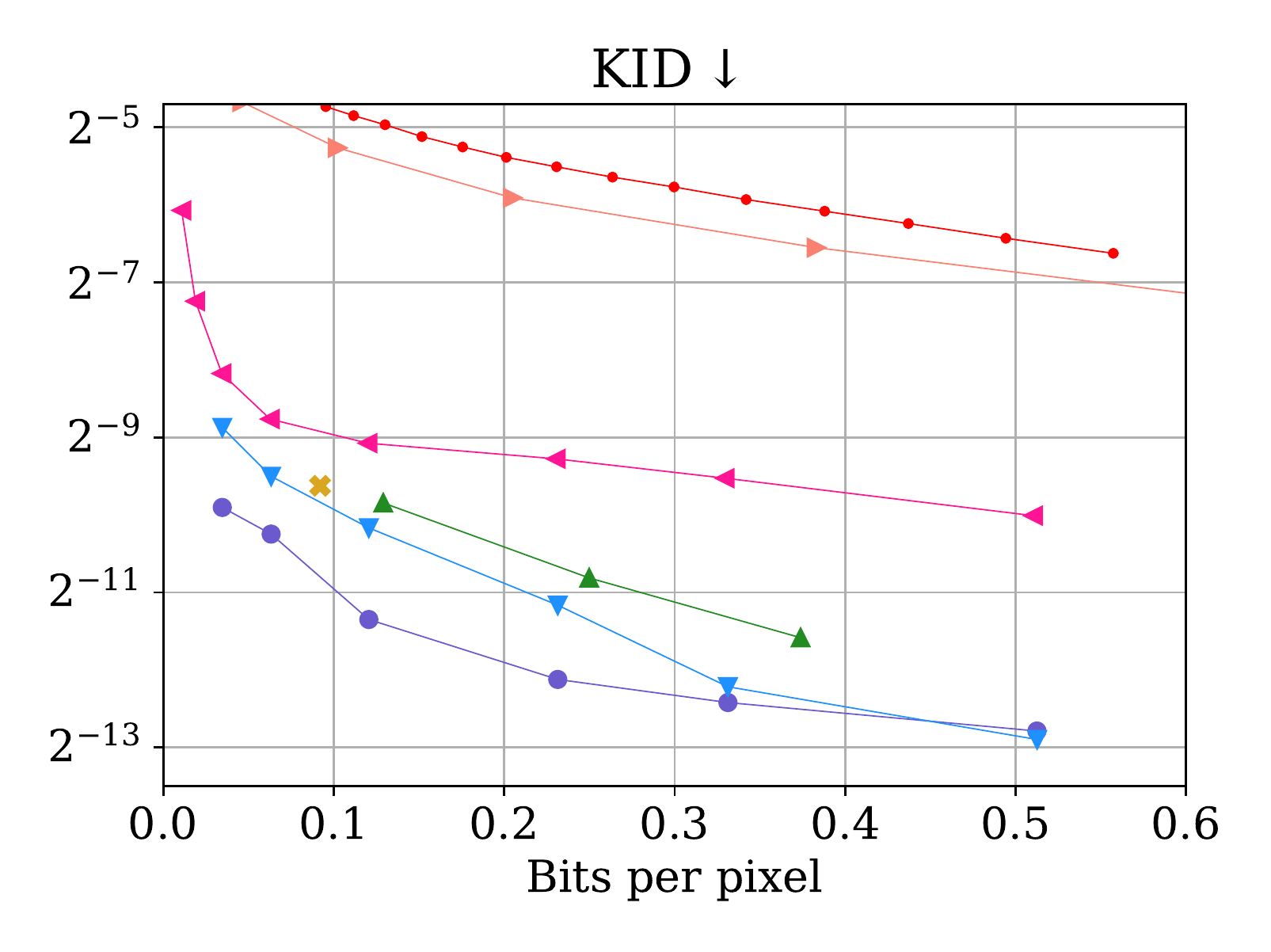}
    \end{subfigure}
    \caption{Comparisons of methods across various distortion and statistical fidelity metrics for the CLIC 2020 test set. Reference models (VTM 19.2, BPG, and No-GAN (Ours)) achieve the best MS-SSIM and PSNR scores, but display poor statistical fidelity as measured by FID and KID. GAN methods (\Ours, \hific, and Multi-Realism) are able to trade some distortion for statistical fidelity. \Ours (Ours) is able to achieve better statistical fidelity as measured by FID and KID vs. HiFiC at equivalent distortion levels.
    }
    \label{fig:rate_plots}
\end{figure*}

In this section we validate our approach with numerical experiments.
We consider metrics that act as surrogates for both reference distortion (i.e., $\rho\left( \recon, \image \right )$) and statistical fidelity (i.e., $d\left (P_{\hat{X}},P_X\right )$) and demonstrate that our approach more efficiently optimizes for distortion and statistical fidelity simultaneously than previous methods.

\subsection{Datasets and metrics}
For training we utilize the train split of OpenImages~V6~\citep{kuznetsova2020open} for all models.
We used the full-resolution versions of the images.
The first augmentation transform is randomly selected to be either a  random resized crop or a simple crop to a standard $256 \times 256$ training resolution.
Our intuition is that this would expose the model to both interpolation statistics as well as raw quantization statistics from standard image codecs.
The second augmentation consists of random horizontal flipping.

For evaluation we adopt the test split of CLIC2020~\cite{todericiclic}, the validation split of DIV2K~\cite{agustsson2017ntire}, and Kodak.\footnote{Kodak {PhotoCD} dataset, \url{http://r0k.us/graphics/kodak}.}
We focus most of our results in the main body on CLIC2020 because it is commonly used by the neural image compression community.
We present results for DIV2K and Kodak in the Appendix. Results for DIV2K and Kodak exhibited similar trends to CLIC2020.

Our evaluation metrics are of two general classes: reference-based and no-reference metrics.
The reference metrics are computed in the form $\rho(\recon,\image)$, where $\image$ is a ground-truth image and $\recon$ is a compressed version of $\image$.
The handcrafted reference metrics of MS-SSIM~\citep{wang2003multiscale} and PSNR are standards for evaluating image compression methods.
A drawback of optimization for the handcrafted metrics is that it can lead to blurring of the reconstructed images.
For this reason, other reference metrics such as LPIPS~\citep{zhang2018unreasonable} and DISTS~\citep{ding2020image} have been developed that more heavily favor preservation of texture and are more correlated with human judgment~\citep{ding2020image}, but it is important to note that as reference metrics they can still trade off some statistical fidelity~\citep{blau2019rethinking}.

Alternatively, no-reference metrics measure statistical fidelity via distributional alignment.
No-reference metrics include FID~\cite{heusel2017gans} and KID~\cite{binkowski2018demystifying}.
Both of these metrics use features from a pretrained Inception V3 model~\citep{szegedy2016rethinking} to parametrize distributional alignment and are used for image generation.

\subsection{Baseline models}

We compare MS-ILLM to a suite of baseline models.
We show details of how we calculated each baseline in the appendix.
\textbf{No-GAN (Ours):} This method is our pretrained model at 1 million steps with no GAN fine-tuning.
\textbf{MS-PatchGAN (HiFiC*):} This the No-GAN model with PatchGAN discriminator fine-tuning, essentially our own training recipe for HiFiC~\citep{mentzer2020high}.
Since we do not have access to the HiFiC training data, this model can be considered as similar to the original, but not an exact replica.
\textbf{VTM 19.2:}\footnote{\url{https://vcgit.hhi.fraunhofer.de/jvet/VVCSoftware_VTM}} The image compression component of VVC, essentially the state-of-the-art handcrafted image compressor.
\textbf{BPG:}\footnote{\url{https://bellard.org/bpg/}} The image compression component of HEVC, adapted to work on images and give minimally-sized headers.
The Mean-Scale Hyperprior autoencoder architecture that we use in our model has similar rate-distortion performance to BPG~\citep{minnen2018joint}.
For this reason, we consider BPG to also be a stand-in for Mean-Scale Hyperprior performance.
\textbf{HiFiC:} The (previous) state-of-the-art generative image compression method of~\citet{mentzer2020high}.
\textbf{Multi-Realism:} A concurrent generative compression method utilizng the recent ELIC autoencoder~\citep{he2022elic} combined with PatchGAN, showing improved performance vs. HiFiC, recently published by~\citet{agustsson2023multi}.
For the multi-realism paper, the authors released a single operating point for CLIC2020, which we include.

To differentiate between in our own training pipeline and that of the original paper, in our plots we use the name ``HiFiC'' only when pulling data from the original paper, and for all other plots we use the name ``MS-PatchGAN (HiFiC*)'' for our own HiFiC training recipe.

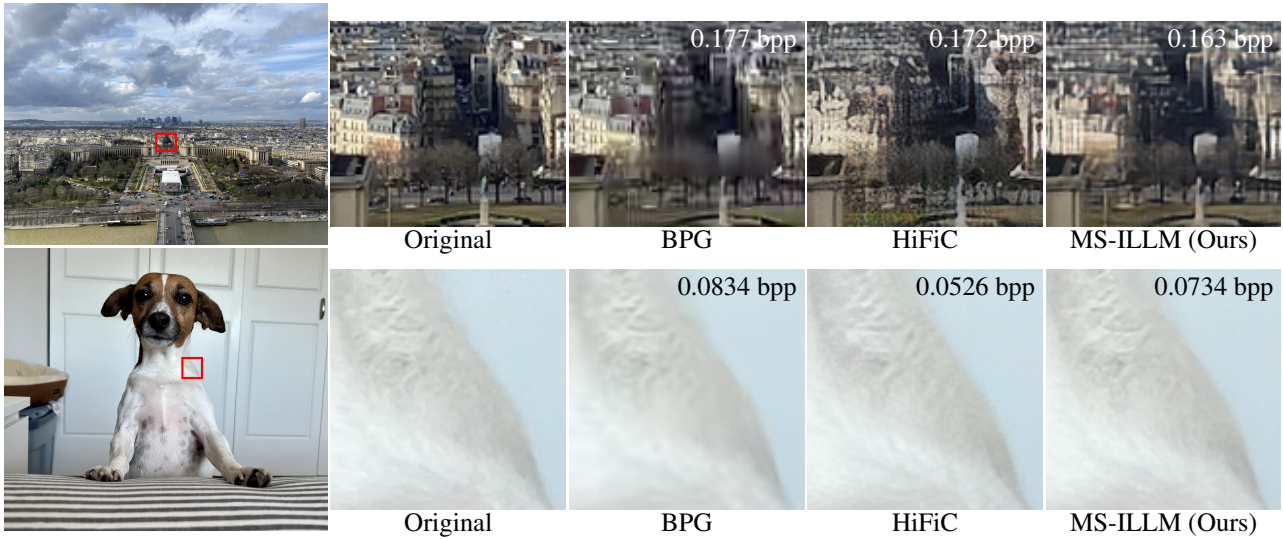
\begin{figure*}[ht]
    \centering
    \begin{tikzpicture}
        \node[anchor=south west,inner sep=0.5pt] (example1) at (0,0)
        {\includegraphics[width=.25\textwidth]{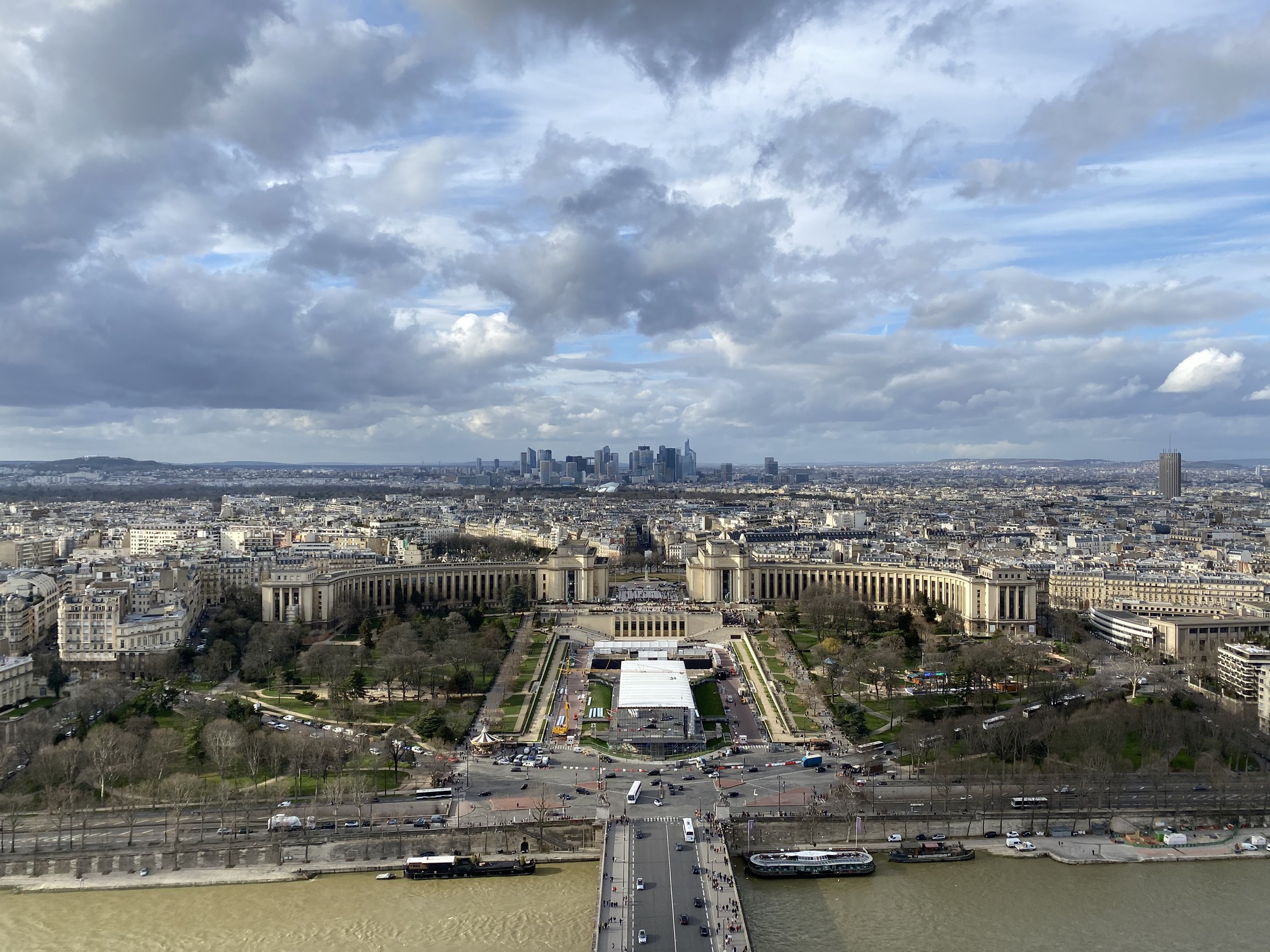}};
        \begin{scope}[x={(example1.south east)},y={(example1.north west)}]
            \draw[red,thick] (0.53,0.39) rectangle (0.47,0.46);
        \end{scope}
        \node[inner sep=0.5pt,anchor=west] (example1crop1orig) at (example1.east)
        {\includegraphics[width=0.182\textwidth]{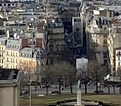}};
        \node[inner sep=0.5pt,anchor=north] (example1crop1origcap) at (example1crop1orig.south) {Original};
        \node[inner sep=0.5pt,anchor=west] (example1crop1bpg) at (example1crop1orig.east)
        {\includegraphics[width=0.182\textwidth]{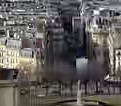}};
        \node[inner sep=0.5pt,anchor=north] (example1crop1bpgcap) at (example1crop1bpg.south) {BPG};
        \node[anchor=north east] at (example1crop1bpg.north east) {\color{white} 0.177 bpp};
        \node[inner sep=0.5pt,anchor=west] (example1crop1hific) at (example1crop1bpg.east)
        {\includegraphics[width=0.182\textwidth]{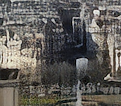}};
        \node[inner sep=0.5pt,anchor=north] (example1crop1hificcap) at (example1crop1hific.south) {HiFiC};
        \node[anchor=north east] at (example1crop1hific.north east) {\color{white} 0.172 bpp};
        \node[inner sep=0.5pt,anchor=west] (example1crop1ours) at (example1crop1hific.east) {\includegraphics[width=0.182\textwidth]{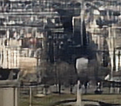}};
        \node[inner sep=0.5pt,anchor=north] (example1crop1ourscap) at (example1crop1ours.south) {\Ours (Ours)};
        \node[anchor=north east] at (example1crop1ours.north east) {\color{white} 0.163 bpp};

        \node[inner sep=0.5pt,anchor=north west] (example2) at (example1.south west) 
        {\includegraphics[width=.25\textwidth]{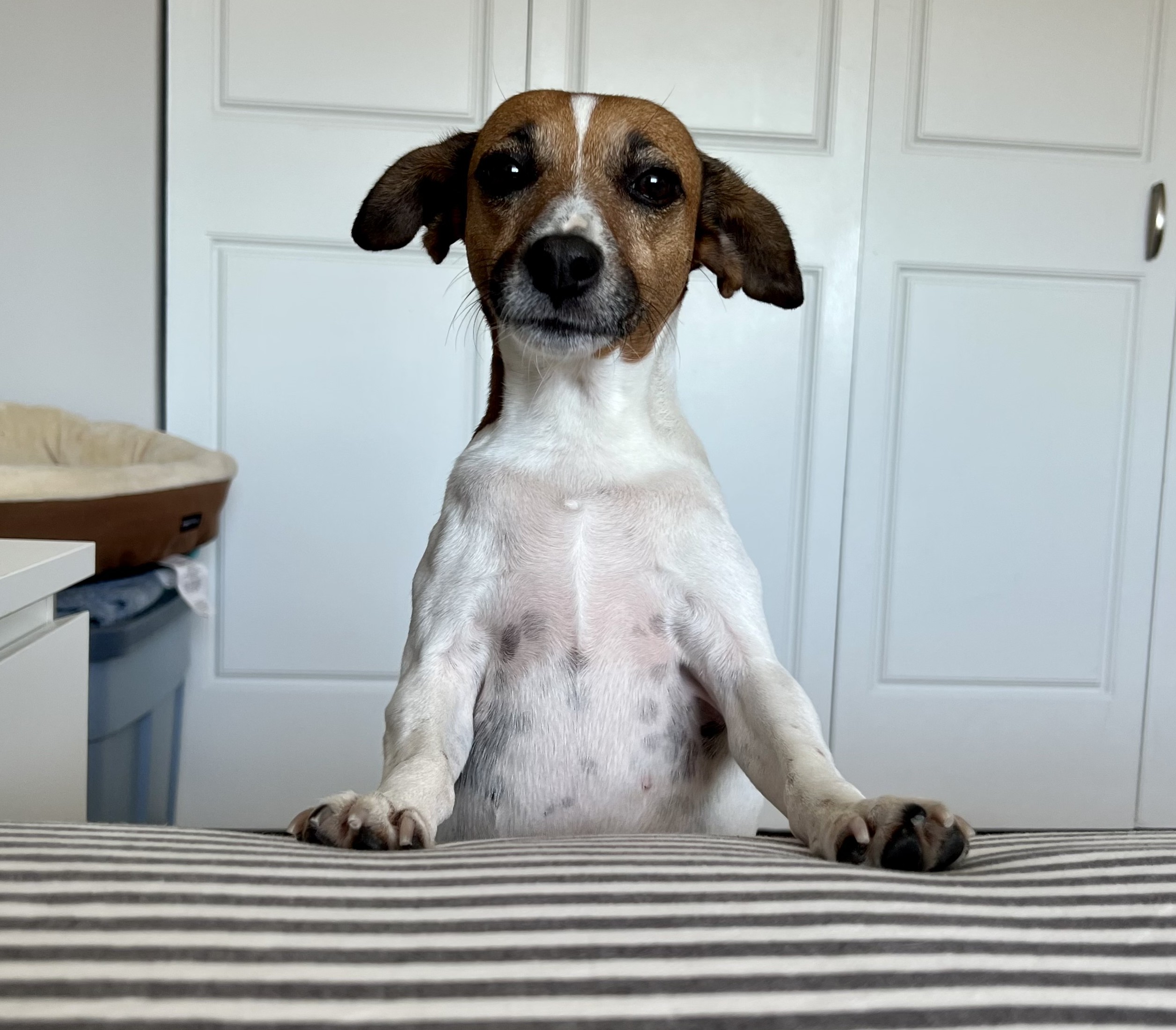}};
        \begin{scope}[x={(example2.north east)},y={(example2.south west)}]
            \draw[red,thick] (0.55,0.39) rectangle (0.61,0.46);
        \end{scope}
        \node[inner sep=0.5pt,anchor=west] (example2crop1orig) at (example2.east)
        {\includegraphics[width=0.182\textwidth]{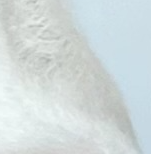}};
        \node[inner sep=0.5pt,anchor=north] (example2crop1origcap) at (example2crop1orig.south) {Original};
        \node[inner sep=0.5pt,anchor=west] (example2crop1bpg) at (example2crop1orig.east)
        {\includegraphics[width=0.182\textwidth]{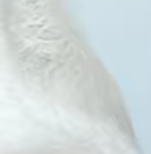}};
        \node[inner sep=0.5pt,anchor=north] (example2crop1bpgcap) at (example2crop1bpg.south) {BPG};
        \node[anchor=north east] at (example2crop1bpg.north east) {\color{black} 0.0834 bpp};
        \node[inner sep=0.5pt,anchor=west] (example2crop1hific) at (example2crop1bpg.east)
        {\includegraphics[width=0.182\textwidth]{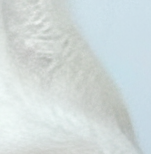}};
        \node[inner sep=0.5pt,anchor=north] (example2crop1hificcap) at (example2crop1hific.south) {HiFiC};
        \node[anchor=north east] at (example2crop1hific.north east) {\color{black} 0.0526 bpp};
        \node[inner sep=0.5pt,anchor=west] (example2crop1ours) at (example2crop1hific.east) {\includegraphics[width=0.182\textwidth]{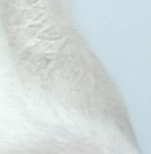}};
        \node[inner sep=0.5pt,anchor=north] (example2crop1ourscap) at (example2crop1ours.south) {\Ours (Ours)};
        \node[anchor=north east] at (example2crop1ours.north east) {\color{black} 0.0734 bpp};
    \end{tikzpicture}
    \caption{Qualitative examples of compressed images. (\textit{top}) A cityscape image from the CLIC2020 test set. For the zoomed insets, BPG shows good quality for the white building in the foreground with quality degrading in the background. HiFiC provides more definition for the background at the cost of compression artifacts. Our method is able to provide the increased definition of HiFiC with fewer compression artifacts. (\textit{bottom}) A custom example with a dog. In this case, BPG leads to some blurring of subtle fur features, as well as blocking compression artifacts. Both blurring and artifacts are removed with both the HiFiC method and ours.
    }
    \label{fig:qualitative-example}
\end{figure*}

\subsection{Main results}
Figure \ref{fig:rate_plots} shows performance of the models across bitrates.
For this figure, we plot using the discriminator weighting $\lambda_d$ in equation (\ref{eq:objective}) such that \Ours, $\lambda_d=0.008$ matches the PSNR performance of \hific from the original paper.
Figure \ref{fig:rate_plots} shows that while matching PSNR, \Ours has uniformly better statistical fidelity than \hific as measured by FID and KID over all bitrates.
Alternatively, with $\lambda_d=0.0005$ we find that we can match HiFiC at a specified FID/KID rate while achieving a higher PSNR.

Figure \ref{fig:teaser} shows the same information from an alternative perspective, where in this case we utilize our own \hific training recipe (\Ourhific) to evaluate at many rate-distortion-perception tradeoff points.
Figure \ref{fig:teaser} shows that for all of the distortion points investigated (distortion in this case being measured by mean-squared error), our ILLM discriminator is able to achieve better statistical fidelity as measured by FID than PatchGAN.

In Figure \ref{fig:qualitative-example} we compare the methods qualitatively in the setting of a cityscape image from CLIC2020 test and a custom pet image.
For the cityscape image, at a bitrate of 0.177 bpp the BPG codec has substantial blurring, most easily observable in the trees.
\hific is able to sharpen the image substantially, but this comes at the cost of introducing distortion artifacts.
Meanwhile, \Ours is able to achieve sharpening with less artifact introduction compared to \hific.
For the pet image, again BPG leads to substantial blurring of subtle textures like fur, while both \hific and \Ours are able to restore much of the missing texture.
We show further qualitative examples with various sharpening levels on the Kodak dataset in the Appendix.

\subsection{Ablations}
\label{sec:ablations}
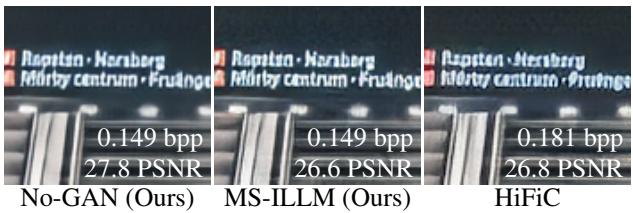
\begin{figure}
    \centering
    \begin{tikzpicture}
        \node[inner sep=0.5pt,anchor=west] (example2crop1nogan) at (0,0)
        {\includegraphics[width=0.16\textwidth]{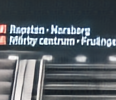}};
        \node[inner sep=0.5pt,anchor=north] (example2crop1nogancap) at (example2crop1nogan.south) {No-GAN (Ours)};
        \node[anchor=south east] (noganpsnr) at (example2crop1nogan.south east) {\color{white} 27.8 PSNR};
        \node[anchor=south east,yshift=-0.15cm] at (noganpsnr.north east) {\color{white} 0.149 bpp};
        \node[inner sep=0.5pt,anchor=west] (example2crop1ours) at (example2crop1nogan.east) {\includegraphics[width=0.16\textwidth]{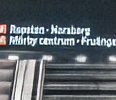}};
        \node[inner sep=0.5pt,anchor=north] (example2crop1ourscap) at (example2crop1ours.south) {\Ours(Ours)};
        \node[anchor=south east] (ourpsnr) at (example2crop1ours.south east) {\color{white} 26.6 PSNR};
        \node[anchor=south east,yshift=-0.15cm] at (ourpsnr.north east) {\color{white} 0.149 bpp};
        \node[inner sep=0.5pt,anchor=west] (example2crop1hific) at (example2crop1ours.east)
        {\includegraphics[width=0.16\textwidth]{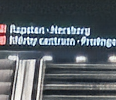}};
        \node[inner sep=0.5pt,anchor=north] (example2crop1hificcap) at (example2crop1hific.south) {HiFiC};
        \node[anchor=south east] (hificpsnr) at (example2crop1hific.south east) {\color{white} 26.8 PSNR};
        \node[anchor=south east,yshift=-0.15cm] at (hificpsnr.north east) {\color{white} 0.181 bpp};
    \end{tikzpicture}
    \caption{An example with text.
    Incorporating adversarial models for both HiFiC and \Ours leads to a degradation in textual quality vs.\ the No-GAN method that only uses reference-based loss functions. Compared to HiFiC, our discriminator is able to slightly improve legibility with the word, ``centrum''.
    \label{fig:text-ablation}}
\end{figure}

In Figure~\ref{fig:text-ablation} we consider a qualitative text example from CLIC2020,  a difficult setting for generative compression models.
In this case we can observe that the No-GAN method is the best at reconstructing text, while both \hific and \Ours lead to text degradation.
As was the case in Figure~\ref{fig:qualitative-example}, \Ours introduces fewer artifacts than \hific.

\begin{table}[t]
\caption{Metrics for different normalization layers.}
\label{tab:normalizationfid}
\vskip 0.15in
\begin{center}
\begin{small}
\begin{sc}
\begin{tabular}{lcccc}
\toprule
 & \multicolumn{2}{c}{None} & \multicolumn{2}{c}{Instance} \\
\cmidrule{2-3} \cmidrule{4-5}
Bitrate & FID$\downarrow$ & PSNR$\uparrow$ & FID$\downarrow$ & PSNR$\uparrow$ \\
\midrule
$\approx$ 0.035 & 6.27 & 27.9 & 8.28 & 28.0 \\
$\approx$ 0.121 & 2.65 & 30.9 & 3.04 & 30.7 \\
$\approx$ 0.231 & 1.77 & 33.0 & 2.21 & 32.8 \\
\bottomrule
\end{tabular}
\end{sc}
\end{small}
\end{center}
\vskip -0.1in
\end{table}
We performed ablations over the normalization layers in our U-Net discriminator.
Normalizations tested included spectral reparametrization~\cite{miyato2018spectral}, instance normalization~\cite{ulyanov2016instance}, and no normalization.
Although spectral normalization was previously demonstrated to work best for PatchGAN in HiFiC~\cite{mentzer2020high} and was also used for OASIS~\cite{sushko2022oasis}, for our setting we found that spectral norm trained extremely slow, taking as much as 25 times the number of gradient steps before it began matching the performance of the other methods.
Surprisingly, we found that no normalization at all was ideal for our setting as demonstrated in Table~\ref{tab:normalizationfid}.
Despite not having normalizations, we did not observe major issues with stability in training.

\begin{table}[t]
\caption{Metrics for different spatial latent size ($H\times W$) and codebook size $C$, for models with bpp $\approx$ 0.121.}
\label{tab:latentsize}
\vskip 0.15in
\begin{center}
\begin{small}
\begin{sc}
\begin{tabular}{cccc}
\toprule
Spatial Size & Codebook Size & FID$\downarrow$ & PSNR$\uparrow$ \\
\midrule
16 $\times$ 16 & \phantom{1}256 & 2.66 & 31.0 \\
16 $\times$ 16 & 1024 & 2.65 & 31.1 \\
32 $\times$ 32 & \phantom{1}256 & 2.57 & 30.9 \\
32 $\times$ 32 & 1024 & 2.55 & 30.8 \\
\bottomrule
\end{tabular}
\end{sc}
\end{small}
\end{center}
\vskip -0.1in
\end{table}
In Table~\ref{tab:latentsize} we examine the effect of latent dimension for the labeling function $u$.
We find that varying the spatial dimension of the latent space has a mild effect on the FID-PSNR tradeoff. The effect of codebook size is almost negligible.

\section{Limitations}
\label{sec:limitations}

The use of adversarial models and neural networks for compression could lead to bias of the output images based on demographic factors such as race or gender.
As such, our results should be considered for research purposes only and not for production systems without further assessment over these factors.
Second, our use of the HiFiC autoencoder requires substantial compute.
Deployment in real-world settings will require miniaturization and heavy quantization of the model weights.
Finally, although we showed improved performance with our method for statistical fidelity as measured by FID/KID, this does not guarantee that our approach would do better in terms of human preference over competing methods. 

\section{Conclusion}
\label{sec:conclusion}
We developed a neural image compression model, called \Ours, that improves  statistical fidelity by using  local adversarial discriminators.
Unlike past discriminators employed in compression, our method emphasizes the locality necessary for the compression task. 
The benefits of this translates into better FID and KID metrics.
As such, we empirically concur with the theory behind the rate-distortion-perception tradeoff for the task of image compression and move the current state of the art closer to the theoretical optimum of perfect statistical fidelity.

\bibliography{egbib}
\bibliographystyle{bibstyle}

\newpage
\appendix
\onecolumn

\renewcommand\thefigure{A\arabic{figure}} 
\setcounter{figure}{0} 

\section*{Improving Statistical Fidelity for Neural Image Compression
with Implicit Local Likelihood Models --- Supplementary Material }

We will upload code for reproducing our results to the NeuralCompression repository at \url{https://github.com/facebookresearch/NeuralCompression}.

\section{Further training details and hyperparameters}
\subsection{Autoencoder pretraining}
For pretraining the autoencoder (i.e., $\encoder$, $\entropy$, and $\decoder$), we used the same overall approach as HiFiC~\citep{mentzer2020high}.
The original paper used adaptive rate targeting by oscillating the value of $\lambda_\rho$ depending on the current empirical rate.
We applied the same approach, where we used a looser $\lambda_\rho$ in early training and increased it after 50,000 training steps.
We trained models with eight different rate targets.
The targets and corresponding values for $\lambda_{\rho,a}$ and $\lambda_{\rho,b}$ are listed in Table \ref{tab:pretrainparams}.
\begin{table}[ht]
    \caption{Hyperparameters for autoencoder pretraining.}
    \label{tab:pretrainparams}
    \vskip 0.15in
    \begin{center}
    \begin{small}
    \begin{sc}
    \begin{tabular}{cccc}
    \toprule
    Target rate & $\lambda_{\rho,a}$ & $\lambda_{\rho,b}$ \\
    \midrule
    0.00875 & $2^5$ & $2^{-4}$ \\
    0.0175 & $2^4$ & $2^{-4}$  \\
    0.035 & $2^3$ & $2^{-4}$  \\
    0.07 & $2^2$ & $2^{-4}$  \\
    0.14 & $2^1$ & $2^{-4}$  \\
    0.30 & $2^0$ & $2^{-4}$  \\
    0.45 & $2^{-1}$ & $2^{-4}$ \\
    0.9 & $2^{-2}$ & $2^{-4}$ \\
    \bottomrule
    \end{tabular}
    \end{sc}
    \end{small}
    \end{center}
    \vskip -0.1in
\end{table}

All models used 1 million steps for pretraining, with the target rate being 1.429 higher for the first 50,000 steps.
Also, the value of $\lambda_{\rho,a}$ was held 50\% lower for the first 50,000 steps.

Complete specification of the cost function also requires specifying $\rho(\hat{x}, x)$.
We utilized a combination of MSE and (AlexNet-based) LPIPS.
All of our images were normalized to the range [0.0, 1.0].
To bring this in line with the original HiFiC implementation, we used the following:
\begin{eqnarray}
    \label{eq:distortionspec}
    \rho(\hat{x}, x) = \lambda_{\text{MSE}} \left \| \hat{x} - x \right \|_2^2 + \text{LPIPS}_{\text{Alex}}\left (\hat{x}, x \right ),
\end{eqnarray}
where $\lambda_{\text{MSE}}=150$.
We applied the same loss at the fine-tuning stage.

\begin{figure}[ht]
    \begin{center}
    \centering
    \begin{subfigure}{0.49\textwidth}
    \includegraphics[width=\textwidth]{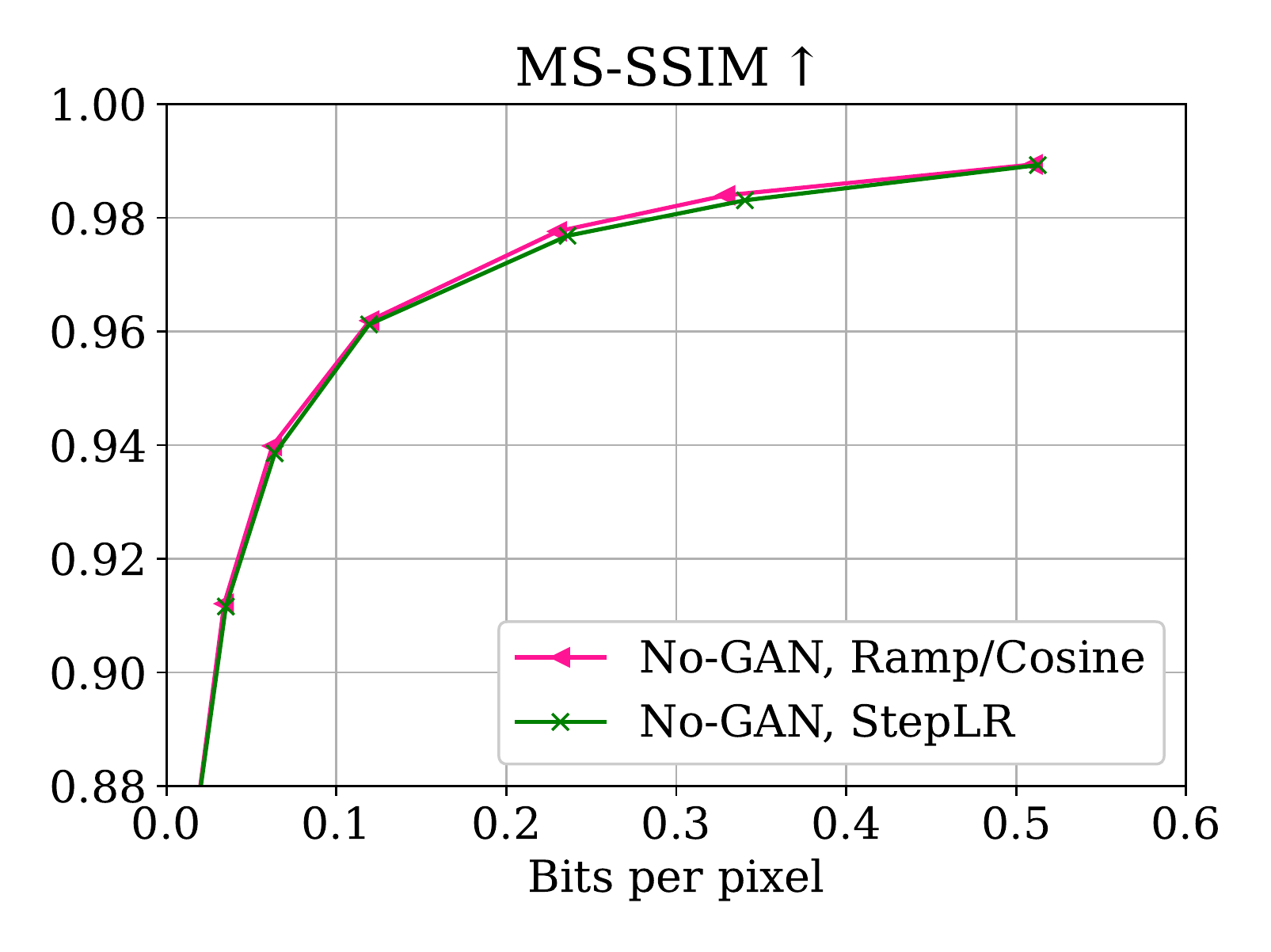}
    \end{subfigure}
    \begin{subfigure}{0.49\textwidth}
    \includegraphics[width=\textwidth]{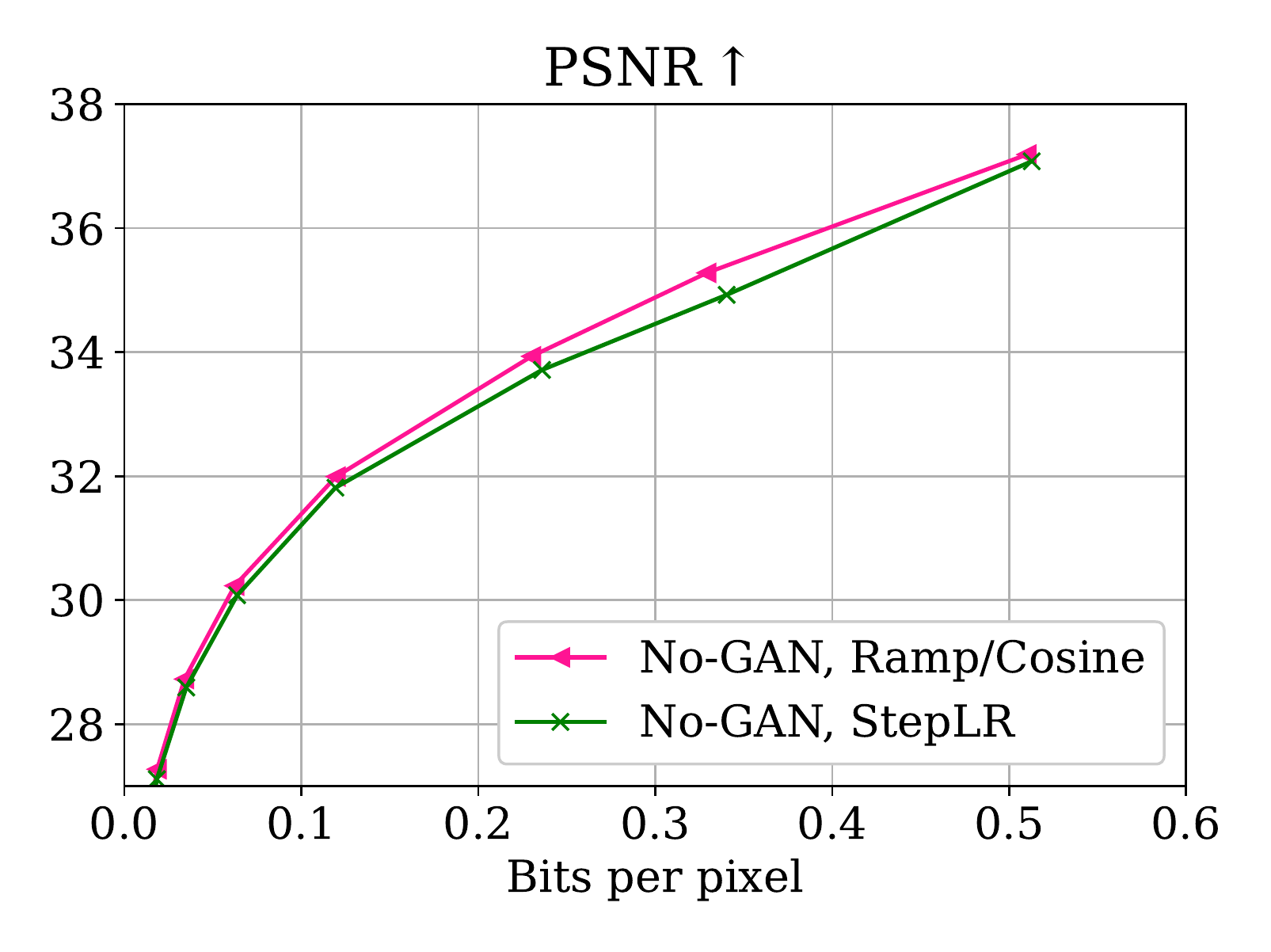}
    \end{subfigure}
    \caption{Comparison of learning rate strategies for training hyperprior models. Ramp/cosine applies a linear warmup for 10,000 steps, followed by cosine decay. The StepLR applies a learning rate decay of 0.1 after 500,000 steps.}
    \label{fig:rdlrcompare}
    \end{center}
\end{figure}

Our last modification for pretraining was to include variable learning rates.
As mentioned in the main body, we observed large gradients at the beginning of training.
At the end of training, very small steps were necessary to acquire the last few dB of PSNR.
The standard approach is to step-decay the LR by a factor of 10 after a significant amount of training (e.g., 500,000 steps).
Our training used a slightly different strategy.
To increase the time the model spent at lower learning rates, we used a linear ramp for the first 10,000 steps followed by cosine rate decay. We observed this gave a small boost to PSNR on the CLIC 2020 test set as shown in Figure \ref{fig:rdlrcompare}.

\subsection{Labeler pretraining}
For pretraining the label function, $\labeler(x)$, we applied the same general approach as \citet{oord2017neural}.
For this case the loss does not include a rate loss, but rather a VQ loss.
To specify the loss, we can introduce the VQ-VAE parameters as $\gamma$ for the encoder, $\zeta$ for the entropy coder (i.e., codebook), and $\psi$ for the decoder.
Then, the loss is
\begin{equation}
    \mathcal{L}(\gamma,\zeta,\psi) = \lambda_{\rho} \text{ } \mathbb{E}_{x \sim P_X}\left [ \rho\left( f_\gamma \circ h_\psi(x),x\right )  \right ] + l_{\text{embedding},\gamma,\zeta}(x) + \beta l_{\text{commitment},\gamma,\zeta}(x),
\end{equation}
where $l_{\text{embedding},\gamma,\zeta}(x)$ and $l_{\text{commitment},\gamma,\zeta}(x)$ are the commitment and embedding losses of the original paper.
For $\rho(\hat{x}, x)$ we used the same equation as in (\ref{eq:distortionspec}), but with $\lambda_{\text{MSE}}=1.0$ and a VGG~\citep{simonyan2014very} backbone instead of AlexNet.

We applied the same learning rate schedule as for the autoencoder pretraining over 1 million training steps.
As the VQ model has a fixed rate, we did not utilize any rate targeting mechanisms.

Despite its role being restricted to labeling for the discriminator, we found that incorporation of perceptual losses in the training of the label function to be critical for its success.
Empirically, we did not observe any good results without the inclusion of LPIPS in $\rho(\hat{x}, x)$ for training the VQ-VAE.

\subsection{Fine-tuning with GAN loss}
For discriminator fine-tuning, we froze both the encoder and the bottleneck of the autoencoder and only fine-tuned the decoder.
For all methods, we utilized a learning rate of 0.0001 for the autoencoder.
For the discriminator, we used a learning rate of 0.0001 for PatchGAN and 0.0004 for ILLM.
For both discriminators, we set the Adam betas to 0.5 and 0.9, following OASIS~\citep{sushko2022oasis}.
We used the non-saturating crossentropy loss for all discriminator training.

\section{Calculation of baseline metrics}
For all methods, we first compressed images with the respective method on each dataset. Then we ran all methods through our own evaluation pipeline to ensure consistency of comparisons.

For VTM we installed VTM 19.2 from the reference at \url{https://vcgit.hhi.fraunhofer.de/jvet/VVCSoftware_VTM}.
First, we converted the images to YCbCr colorspace using the \texttt{rgb2ycbcr} command from \texttt{compressai}~\citep{begaint2020compressai}.
Then, we ran the following:
\begin{verbatim}
# Encode
$VTM_DIR/bin/EncoderAppStatic \
    -i $INPUT_YUV \
    -c $VTM_DIR/cfg/encoder_intra_vtm.cfg \
    -q $QUALITY \
    -o /dev/null \
    -b $COMPRESSED_FILE \
    -wdt $WIDTH \
    -hgt $HEIGHT \
    -fr 1 \
    -f 1 \
    --InputChromaFromat=444 \
    --InputBitDepth=8 \
    --ConformanceWindowMode=1

# Decode
$VTM_DIR/bin/DecoderAppStatic \
    -b $COMPRESSED_FILE \
    -o $OUTPUT_YUV \
    -d 8
\end{verbatim}

For BPG, we used
\begin{verbatim}
# encode
bpgenc -q $QUALITY $INPUT_IMAGE -o $COMPRESSED_FILE -m 9 -f 444

# decode
bpgdec -o $OUTPUT_IMAGE $COMPRESSED_FILE
\end{verbatim}

For HiFiC~\citep{mentzer2020high}, we used the \texttt{tfci} command from Tensorflow Compression~\citep{tfc_github}.

For the multirealism results~\citep{agustsson2023multi}, we used the images uploaded at \url{https://storage.googleapis.com/multi-realism-paper/multi_realism_paper_supplement.zip}.

\section{Note on metrics computation: bitrate, FID, and KID}

In order to decode, our entropy coder $g_{\omega}$ and image decoder $h_{\upsilon}$ require both the bitstreams and metadata that specify the size of the image.
To consider this for the bitrate, we wrote the bistreams and metadata to \texttt{pickle} files and measured the size of the file to estimate bitrates for all methods.
For the hyperprior methods and HiFiC, we measured the size of the \texttt{tfci} file output by \texttt{tensorflow-compression} to estimate their rates.
For the codec methods (BPG and VTM), we measured the size of the file to measure its rate.
BPG and VTM are fully-featured and contain some extra metadata vs. the other methods, so this leads to a slight overestimate of the rate.
However, this is a small cost overhead per image and does not majorly impact the results we present.

For the calculation of FID and KID, we used the \texttt{torch-fidelity} package (available from \url{https://github.com/toshas/torch-fidelity}) to calculate all metrics on our paper.
Previous methods have used \texttt{tensorflow} (e.g., \texttt{tensorflow-gan}), but \texttt{torch-fidelity} has a few differences from \texttt{tensorflow-gan}:
\begin{enumerate}
    \item \texttt{torch-fidelity} includes a standardized Inception V3 module~\citep{szegedy2016rethinking} for feature extraction that includes Tensorflow 1.0-compatible resizing of the input images to a 299 $\times$ 299 resolution, whereas \texttt{tensorflow-gan} leaves the image resizing up to the user.
    \item For the calculation of KID, \texttt{torch-fidelity} bootstraps the estimates by sampling subsets with replacement, whereas \texttt{tensorflow-gan} evenly divides the features into equal subsets with a maximum number of elements per subset.
\end{enumerate}
After adjusting for these changes, we were able to match the two package implementations up to machine precision, but despite this we were unable to match the exact numbers from previous papers.
For example, HiFiC~\citep{mentzer2020high} reports a 4.19736528 FID for HiFiC-Lo, whereas our implementation finds an FID of 4.18498420715332 for the released images.
The differences for KID are even larger.

To provide a fair comparison, we opt instead to recalculate FID and KID for all baseline models in our work with the \texttt{torch-fidelity} package, meaning that in our plots we use the more optimistic FID for HiFiC of $\approx$ 4.185 rather than the pessimistic FID of $\approx$ 4.197.
We also use the KID implementation of \texttt{torch-fidelity} to match those implemented by other groups in the PyTorch community.

\section{Further analysis of the rate-distortion-perception tradeoff}
\begin{figure}[ht]
    \begin{center}
    \centering
    \begin{subfigure}{0.49\textwidth}
    \includegraphics[width=\textwidth]{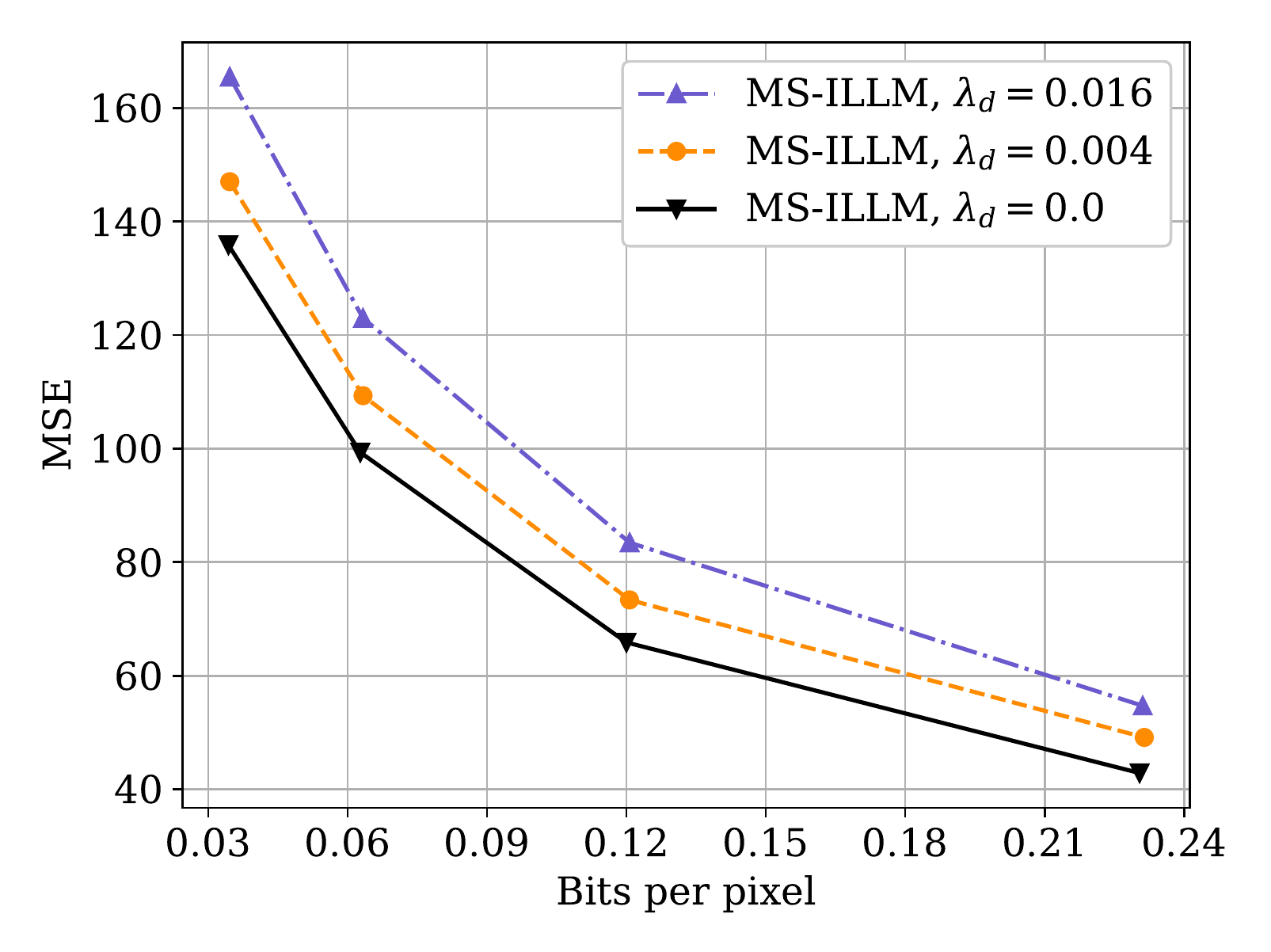}
    \end{subfigure}
    \begin{subfigure}{0.49\textwidth}
    \includegraphics[width=\textwidth]{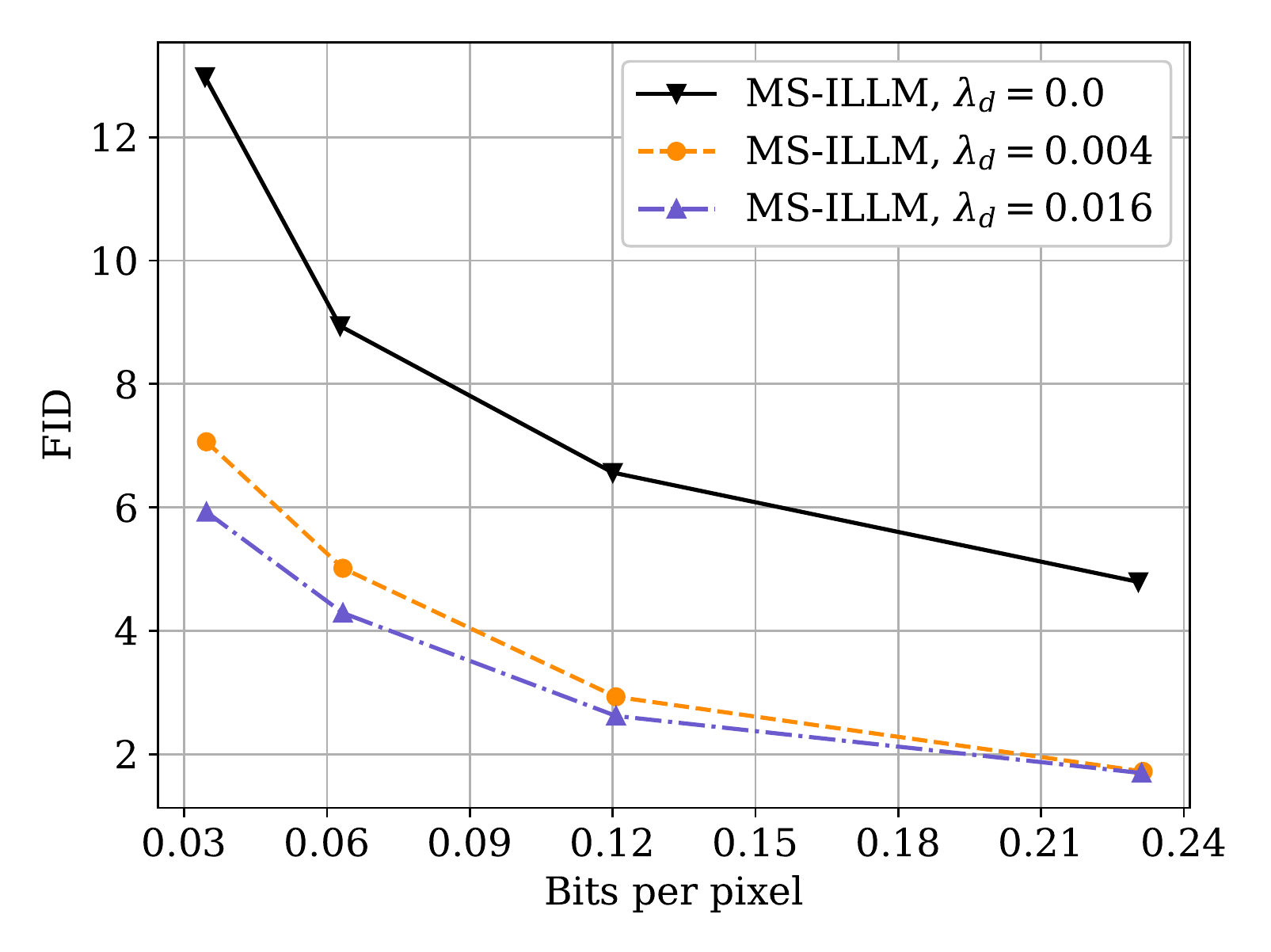}
    \end{subfigure}
    \caption{(\textit{left}) Empirical validation of Figure 1 from \cite{blau2019rethinking} for the task of image compression. 
    As the weight for statistical fidelity, $\lambda_d$ is increased, we find that distortion (as measured by MSE) deteriorates, while statistical fidelity as measured by FID improves (\textit{right}).
    }
    \label{fig:rate-distortion-fixlam}
    \end{center}
\end{figure}

Figure \ref{fig:rate-distortion-fixlam} shows rate-distortion curves over several weight settings $\lambda_d$ for eq.\ (\ref{eq:RDP-fun}) when training the generator.
As the weight increases, statistical alignment (as measured by FID) between compressed and real images improves while distortion suffers, confirming the theoretical results of~\citet{blau2019rethinking}.

\section{Additional qualitative examples}
\begin{figure*}[ht]
    \centering
    \begin{tikzpicture}
        \node[inner sep=0.5pt,anchor=west] (example2crop1orig) at (0,0)
        {\includegraphics[width=0.23\textwidth]{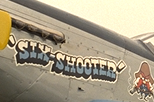}};
        \node[inner sep=0.5pt,anchor=north] (example2crop1origcap) at (example2crop1orig.south) {Original};
        \node[inner sep=0.5pt,anchor=west] (example2crop1hific) at (example2crop1orig.east)
        {\includegraphics[width=0.23\textwidth]{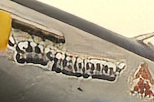}};
        \node[inner sep=0.5pt,anchor=north] (example2crop1hificcap) at (example2crop1hific.south) {HiFiC};
        \node[anchor=north east] at (example2crop1hific.north east) {\color{black} 0.12 bpp};
        \node[inner sep=0.5pt,anchor=west] (example2crop1ours) at (example2crop1hific.east) {\includegraphics[width=0.23\textwidth]{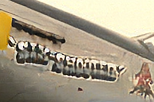}};
        \node[inner sep=0.5pt,anchor=north] (example2crop1ourscap) at (example2crop1ours.south) {\Ours, $\lambda_d=0.008$};
        \node[anchor=north east] at (example2crop1ours.north east) {\color{black} 0.092 bpp};
        \node[inner sep=0.5pt,anchor=west] (example2crop1ourspsnr) at (example2crop1ours.east) {\includegraphics[width=0.23\textwidth]{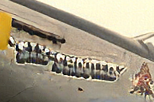}};
        \node[inner sep=0.5pt,anchor=north] (example2crop1ourspsnrcap) at (example2crop1ourspsnr.south) {\Ours, $\lambda_d=0.016$};
        \node[anchor=north east] at (example2crop1ourspsnr.north east) {\color{black} 0.092 bpp};

        \node[inner sep=0.5pt,anchor=west] (example1crop1orig) at (example2.south west)
        {\includegraphics[width=0.23\textwidth]{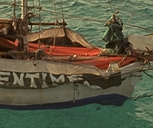}};
        \node[inner sep=0.5pt,anchor=north] (example1crop1origcap) at (example1crop1orig.south) {Original};
        \node[inner sep=0.5pt,anchor=west] (example1crop1hific) at (example1crop1orig.east)
        {\includegraphics[width=0.23\textwidth]{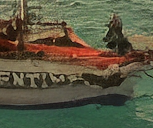}};
        \node[inner sep=0.5pt,anchor=north] (example1crop1hificcap) at (example1crop1hific.south) {HiFiC};
        \node[anchor=north east] at (example1crop1hific.north east) {\color{white} 0.196 bpp};
        \node[inner sep=0.5pt,anchor=west] (example1crop1ours) at (example1crop1hific.east) {\includegraphics[width=0.23\textwidth]{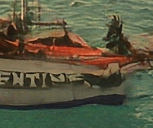}};
        \node[inner sep=0.5pt,anchor=north] (example1crop1ourscap) at (example1crop1ours.south) {\Ours, $\lambda_d=0.008$};
        \node[anchor=north east] at (example1crop1ours.north east) {\color{white} 0.155 bpp};
        \node[inner sep=0.5pt,anchor=west] (example1crop1ourspsnr) at (example1crop1ours.east) {\includegraphics[width=0.23\textwidth]{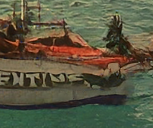}};
        \node[inner sep=0.5pt,anchor=north] (example1crop1ourspsnrcap) at (example1crop1ourspsnr.south) {\Ours, $\lambda_d=0.016$};
        \node[anchor=north east] at (example1crop1ourspsnr.north east) {\color{white} 0.155 bpp};
    \end{tikzpicture}
    \caption{Qualitative examples with MS-ILLM compared to HiFiC on images from the Kodak dataset. The $\lambda_d=0.008$ result shows MS-ILLM with approximately the same PSNR as HiFiC as computed on the CLIC 2020 test dataset. This model is operating at a lower FID than HiFiC, and displays fewer artifacts in the examples. By increasing $\lambda_d$ MS-ILLM can more gracefully add details and textures to the image without increasing artifacts compared to the PatchGAN method of HiFiC.}
    \label{fig:app-qualitative-example}
\end{figure*}

In Figure~\ref{fig:app-qualitative-example} we show the effect of more aggressive discriminator weighting using the ILLM discriminator vs. PatchGAN (as used in HiFiC). When the two methods are matched for PSNR, ILLM has a better FID (see Figure~\ref{fig:rate_plots}), and shows fewer artifacts. As we increase the discriminator weight, ILLM allows even more sharpening and detail addition without artifact reduction.

\section{Experimental results on DIV2K and Kodak}
\begin{figure*}[ht]
    \centering
    \begin{subfigure}{0.8\textwidth}
        \includegraphics[width=\textwidth]{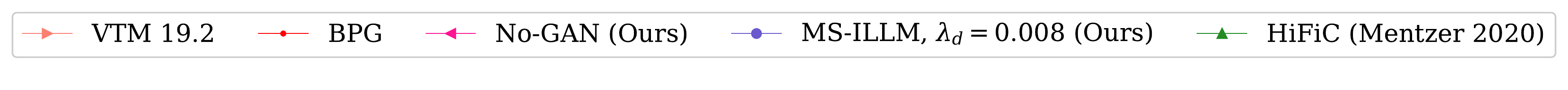}
    \end{subfigure}
    
    \begin{subfigure}{.33\textwidth}
        \includegraphics[width=\textwidth]{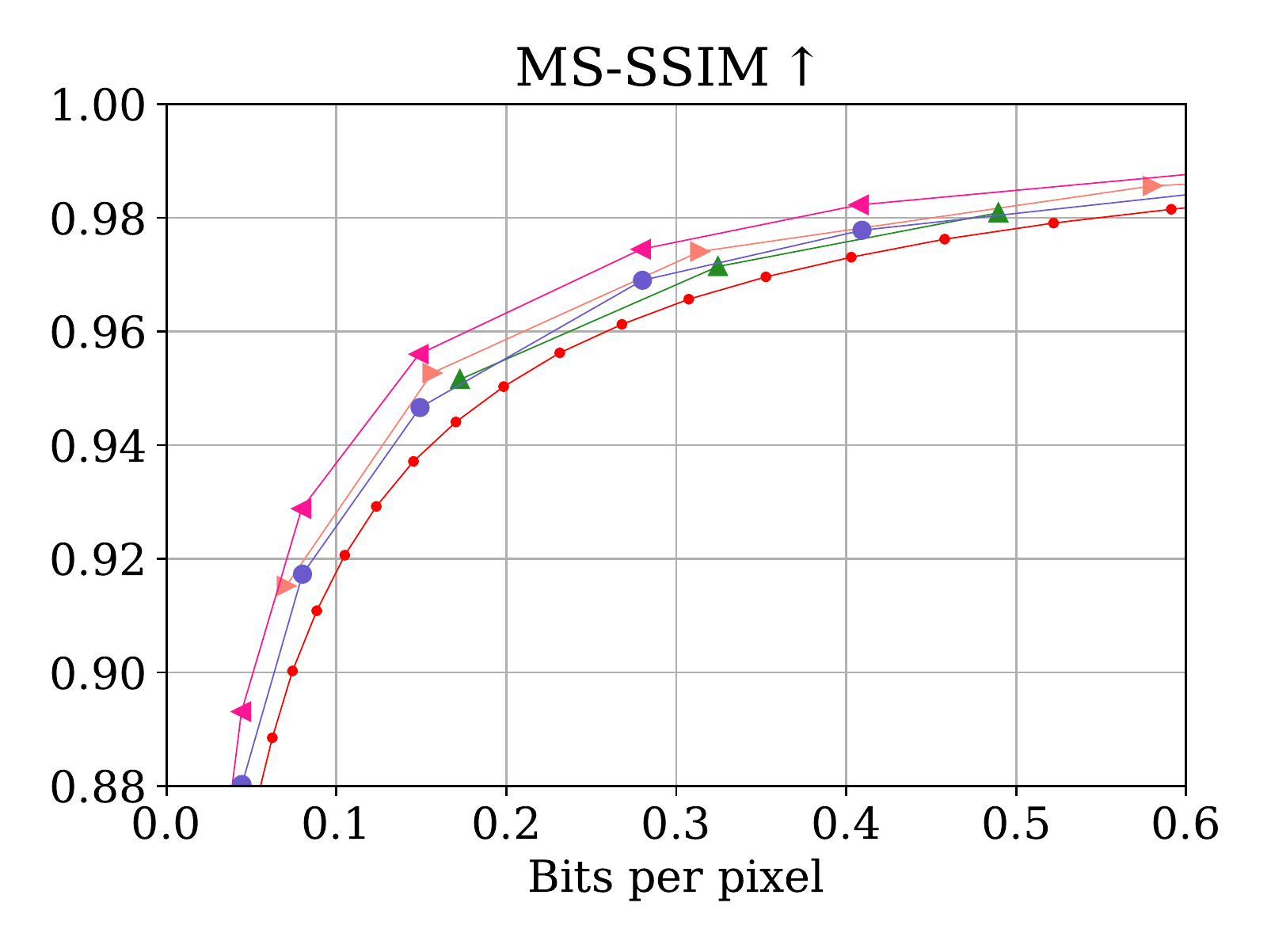}
    \end{subfigure}
    \begin{subfigure}{.33\textwidth}
        \includegraphics[width=\textwidth]{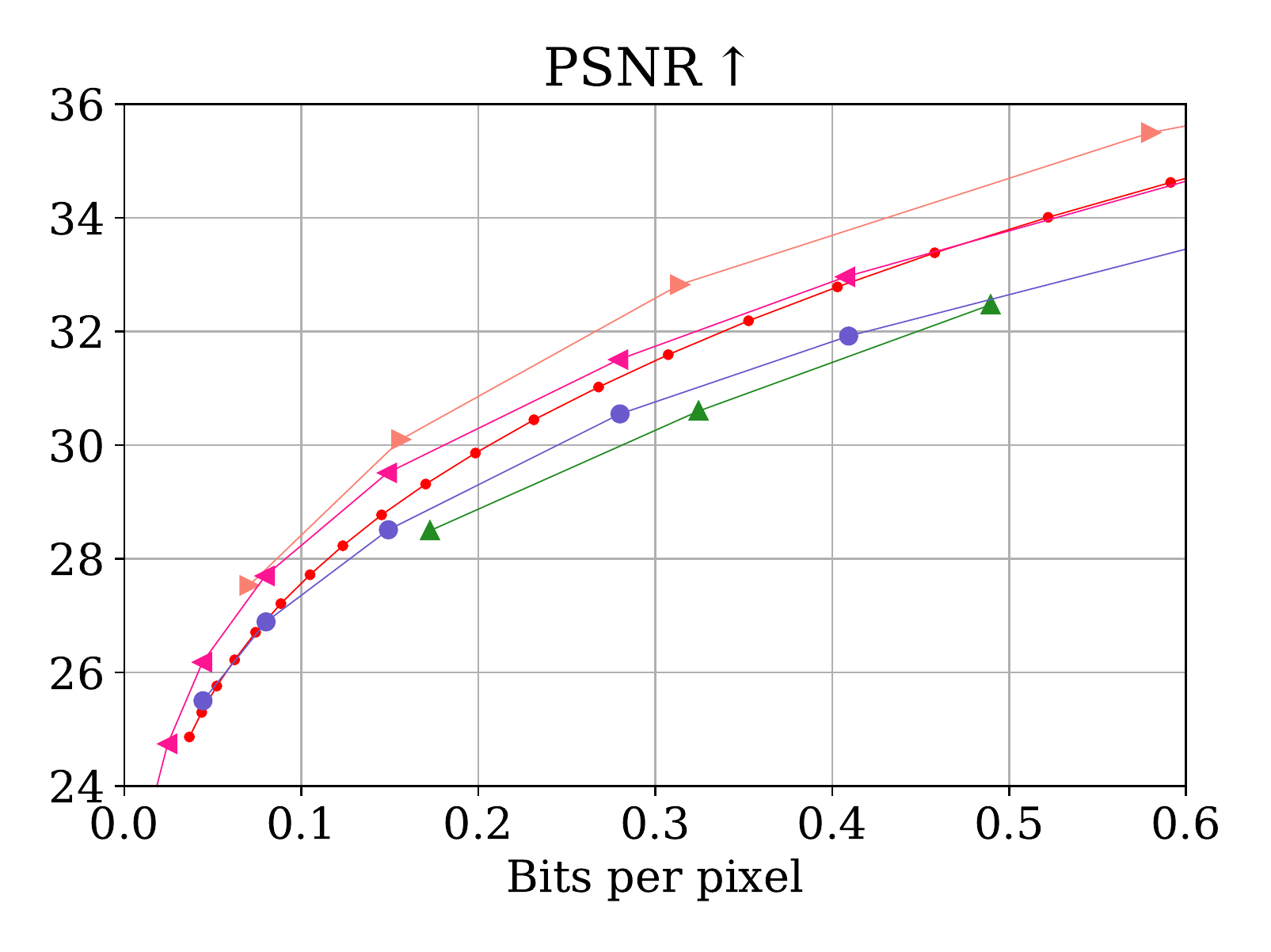}
    \end{subfigure}
    \begin{subfigure}{.33\textwidth}
        \includegraphics[width=\textwidth]{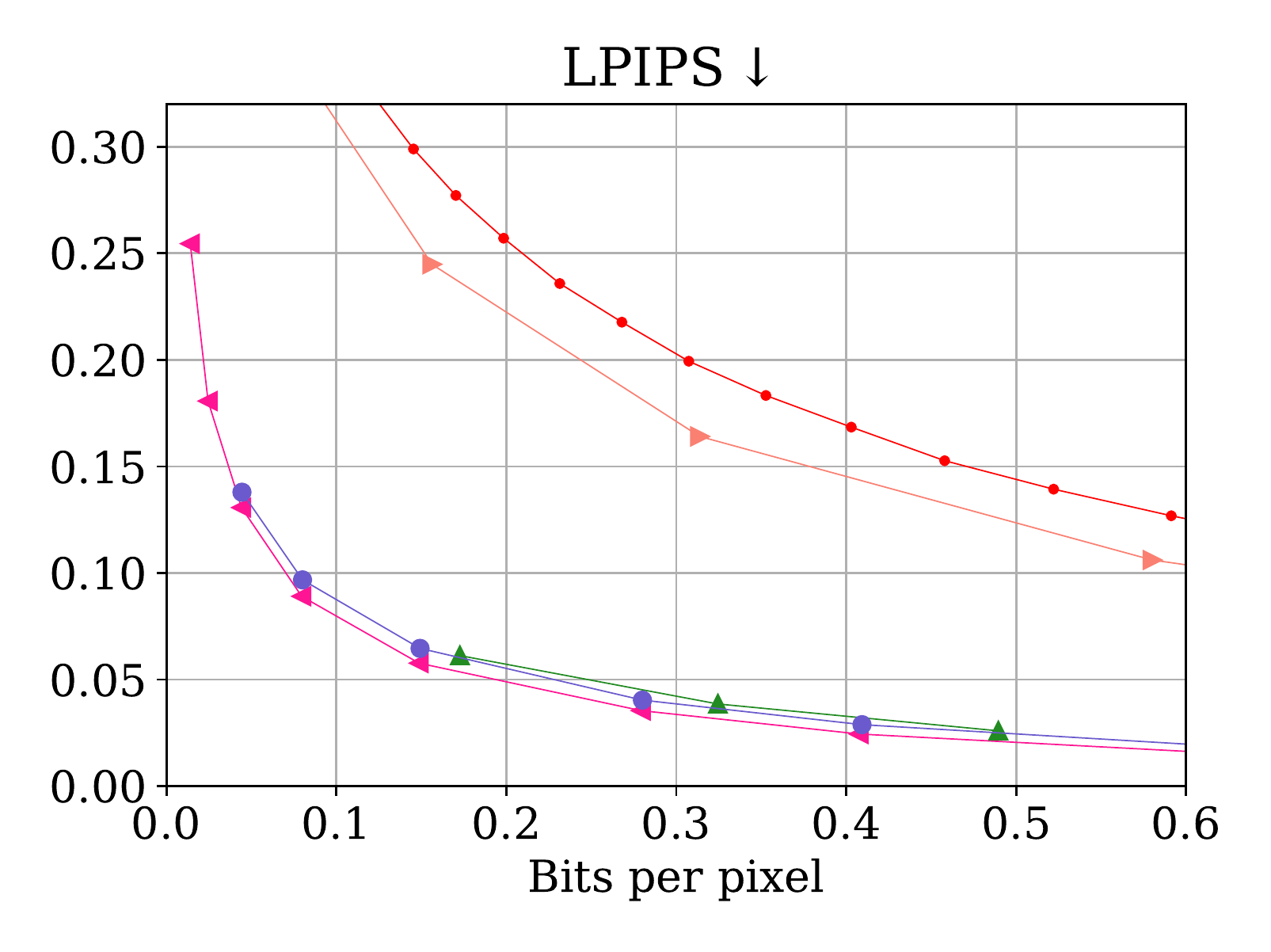}
    \end{subfigure}
    \begin{subfigure}{.33\textwidth}
        \includegraphics[width=\textwidth]{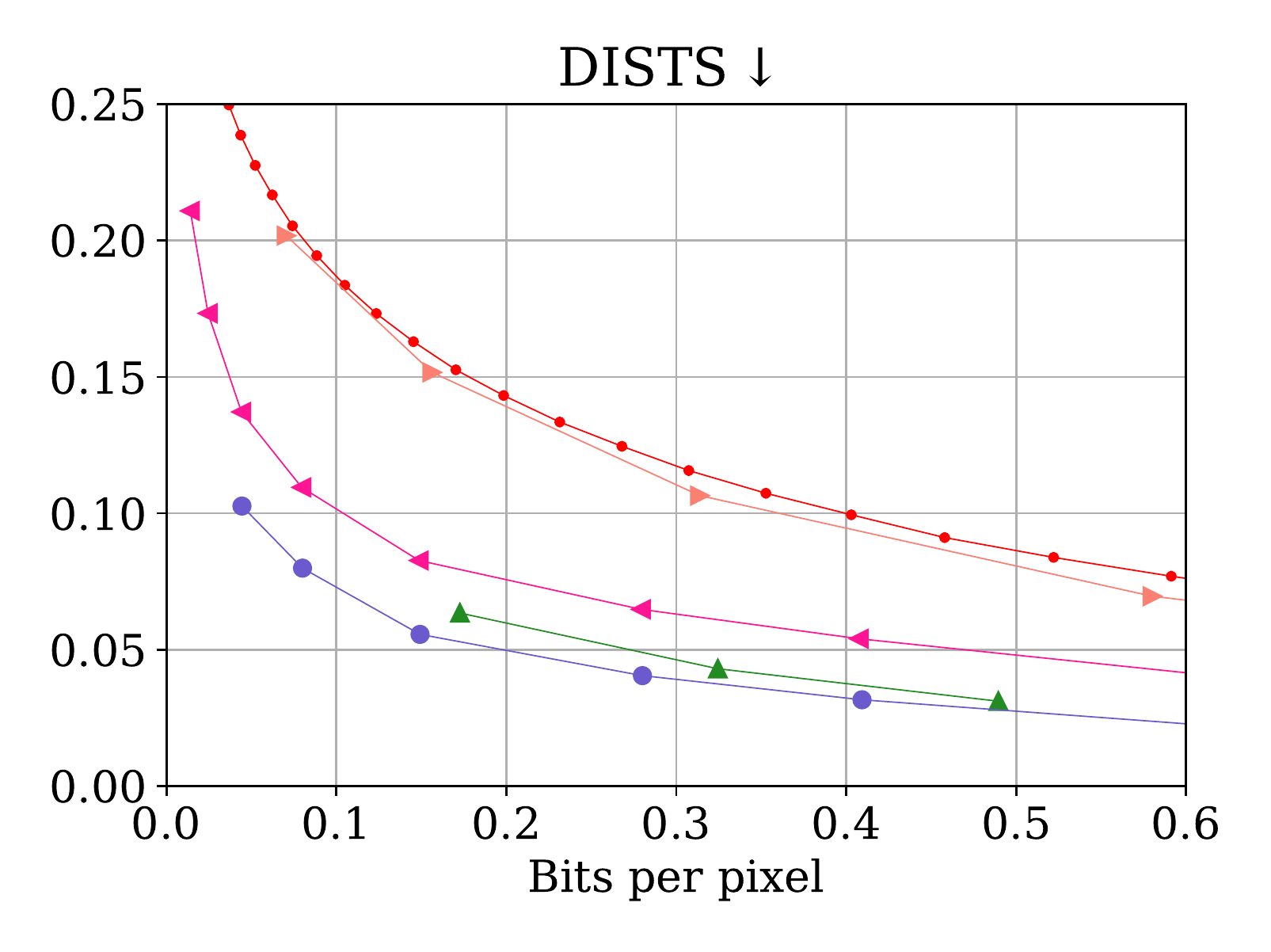}
    \end{subfigure}
    \begin{subfigure}{.33\textwidth}
        \includegraphics[width=\textwidth]{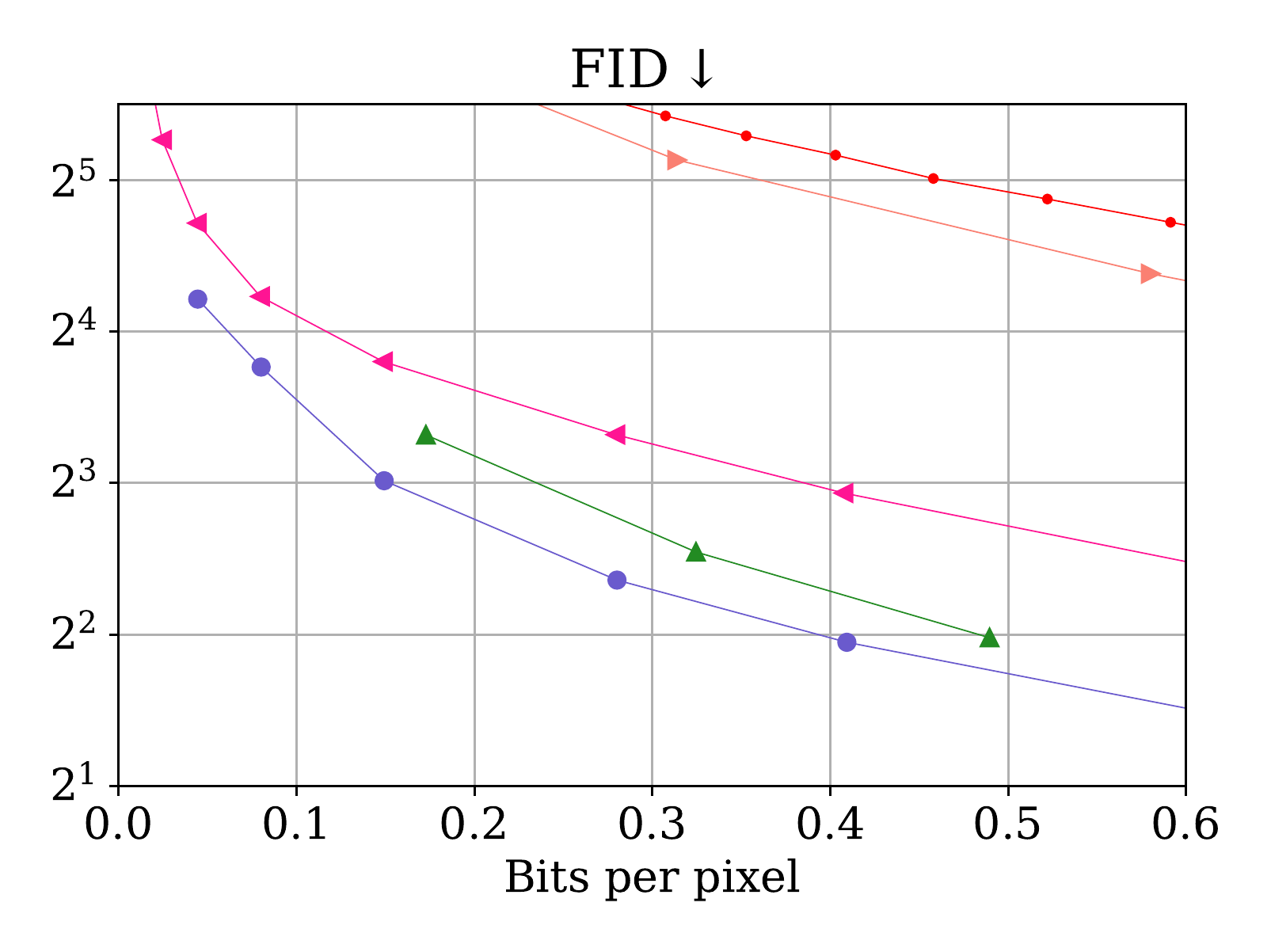}
    \end{subfigure}
    \begin{subfigure}{.33\textwidth}
        \includegraphics[width=\textwidth]{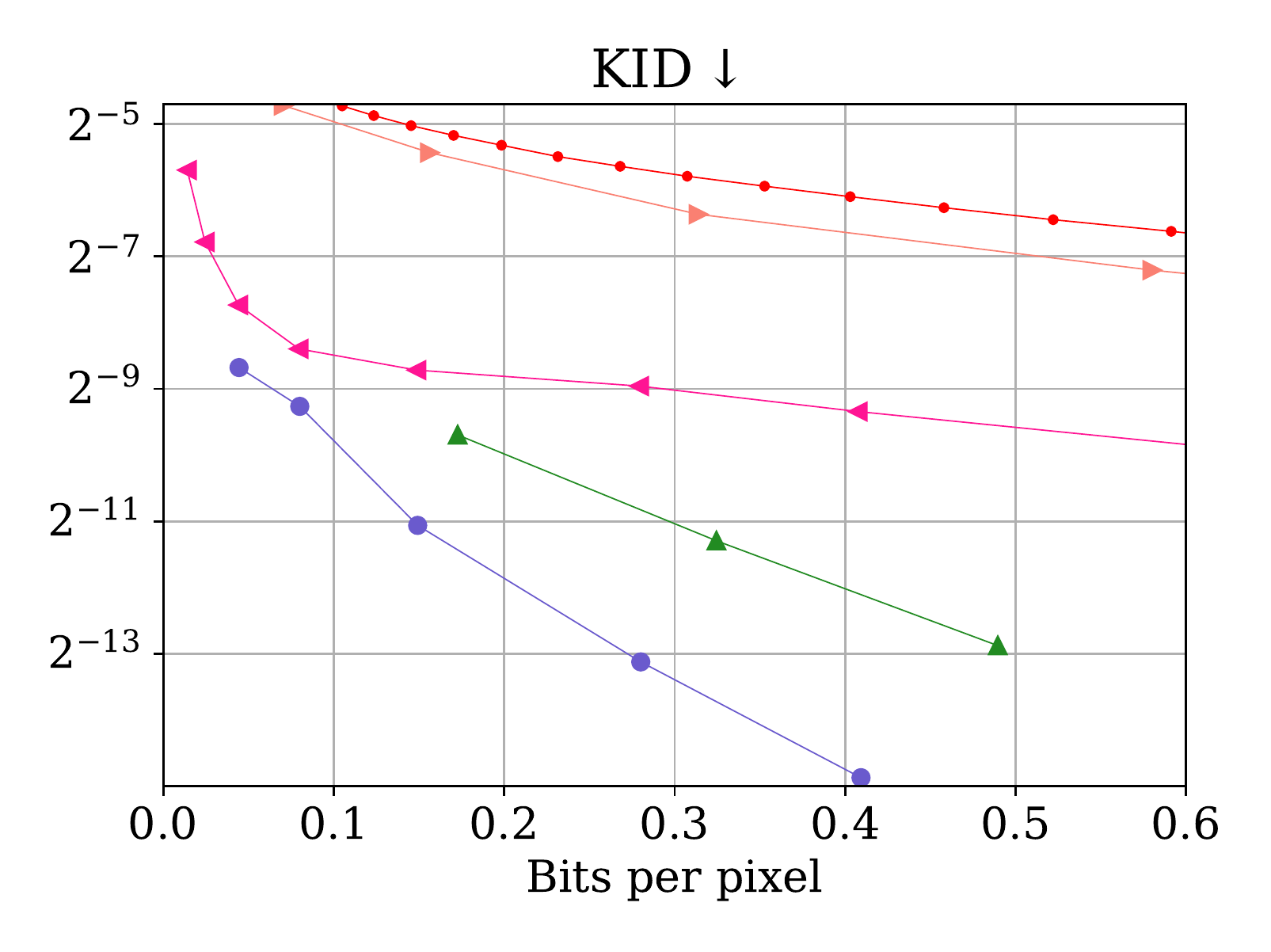}
    \end{subfigure}
    \caption{Comparisons of methods across various distortion and statistical fidelity metrics for the DIV2K validation set. As with Figure~\ref{fig:rate_plots} in the main body, MS-ILLM is able to match HiFiC in reference metrics (MS-SSIM, PSNR, LPIPS, and DISTS) while outperforming HIFiC in no-reference metrics (FID and KID) that indicate statistical fidelity.
    }
    \label{fig:div2k_rate_plots}
\end{figure*}

We provide further experimental results on the DIV2K validation set~\citep{agustsson2017ntire}, shown in Figure~\ref{fig:div2k_rate_plots}.
As with Figure~\ref{fig:rate_plots} in the main body, \Ours is able to match HiFiC on reference-based metrics (MS-SSIM, PSNR, LPIPS, DISTS) while outperforming HiFiC on no-reference distributional metrics (FID and KID).
This further supports the claim that our discriminator is more efficient in trading distortion for statistical fidelity.

We note that for DIV2K we observed some instabilities in calculating KID.
These were small enough that it was not an issue for CLIC2020 test or DIV2K at lower bitrates.
However, at higher bitrates we observed that it was possible for our method to yield negative KID scores.
For this reason, in Figure~\ref{fig:div2k_rate_plots} we do not plot the last point of KID for our method.

\begin{figure*}[ht]
    \centering
    \begin{subfigure}{0.8\textwidth}
        \includegraphics[width=\textwidth]{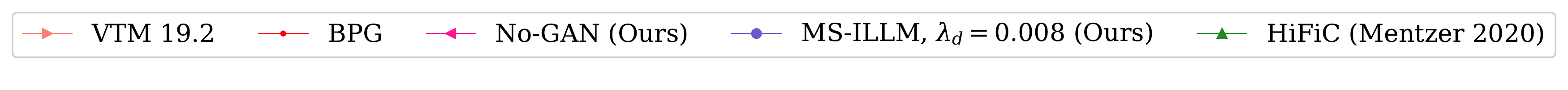}
    \end{subfigure}
    
    \begin{subfigure}{.33\textwidth}
        \includegraphics[width=\textwidth]{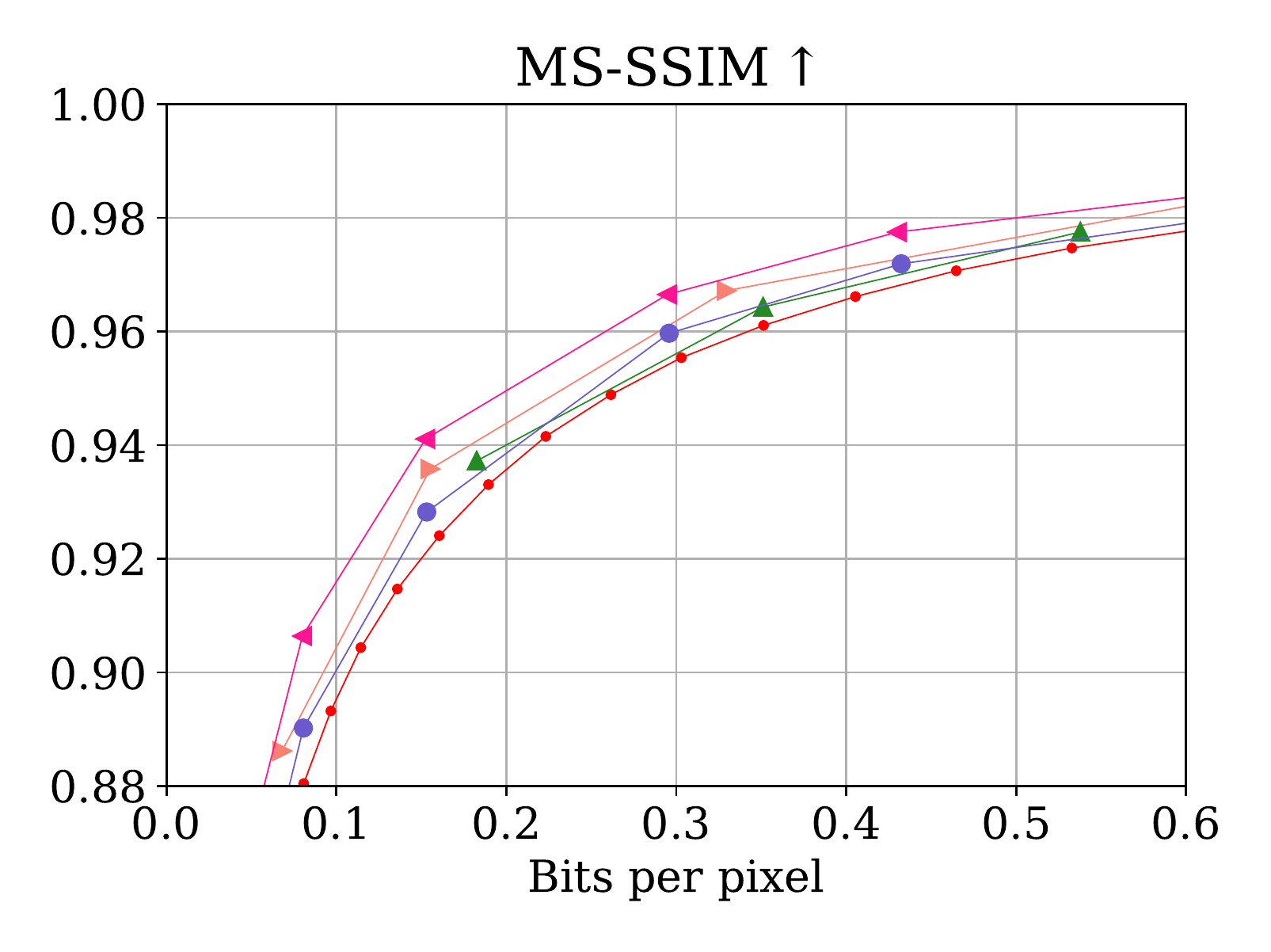}
    \end{subfigure}
    \begin{subfigure}{.33\textwidth}
        \includegraphics[width=\textwidth]{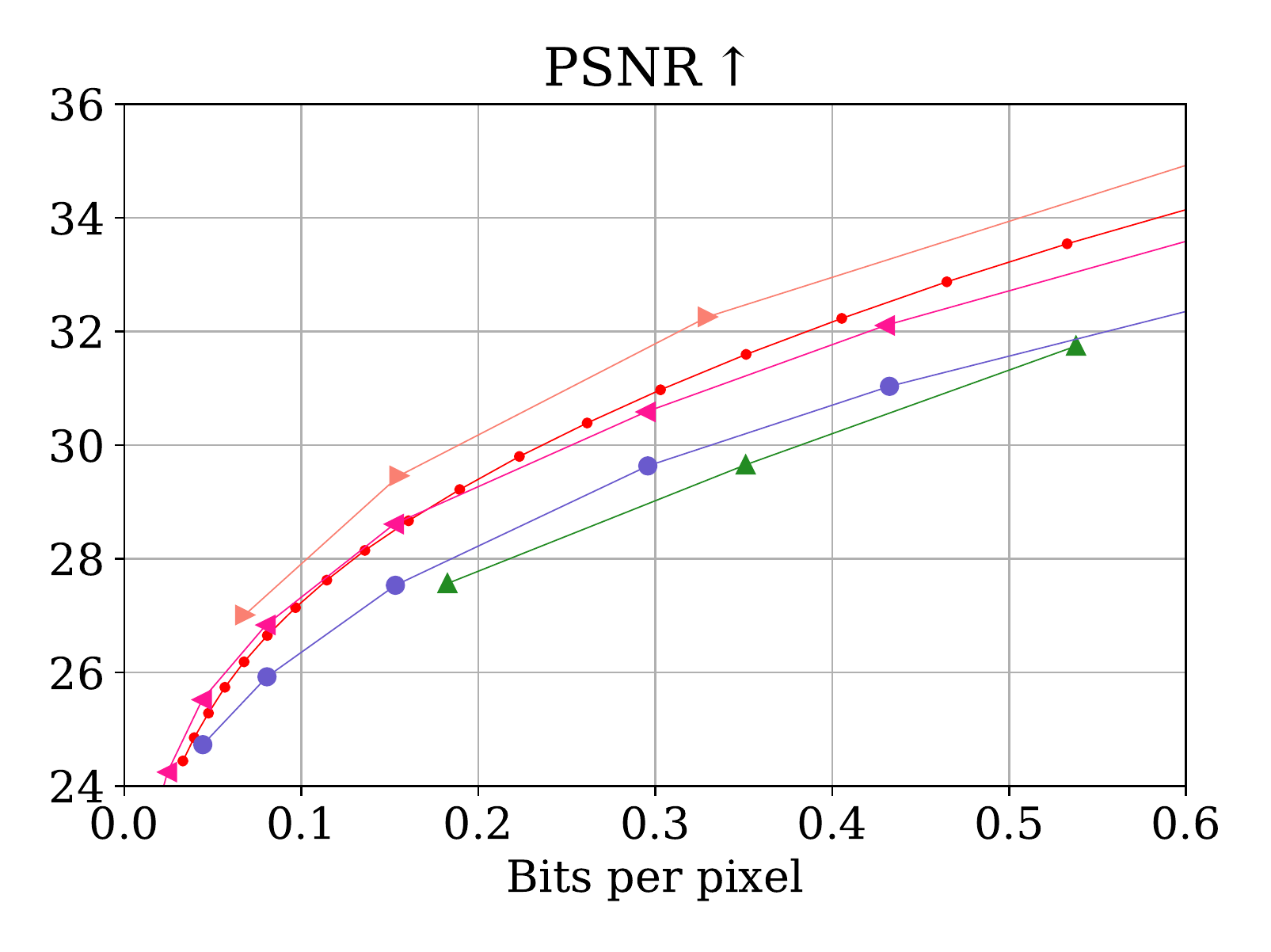}
    \end{subfigure}

    \begin{subfigure}{.33\textwidth}
        \includegraphics[width=\textwidth]{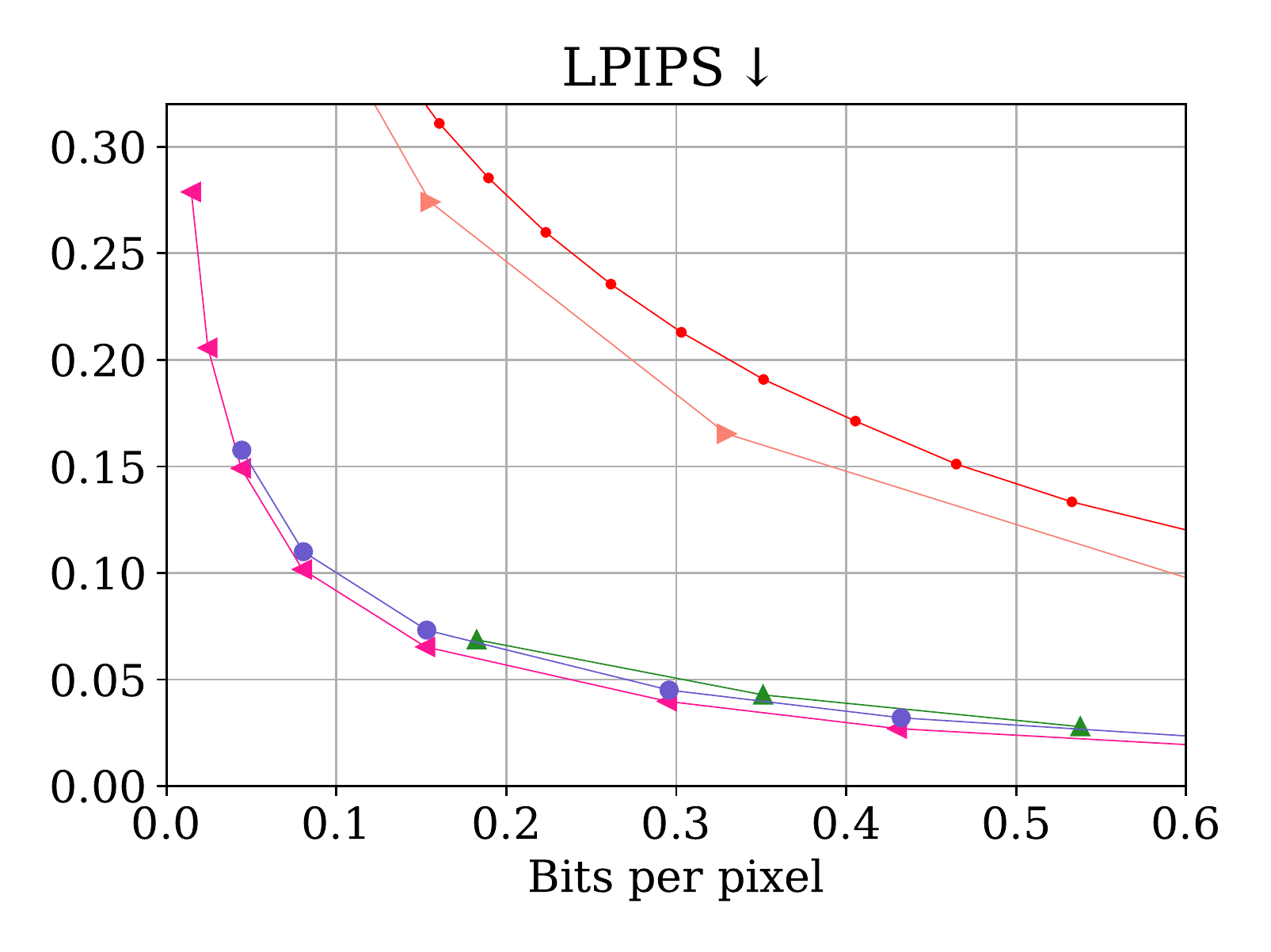}
    \end{subfigure}
    \begin{subfigure}{.33\textwidth}
        \includegraphics[width=\textwidth]{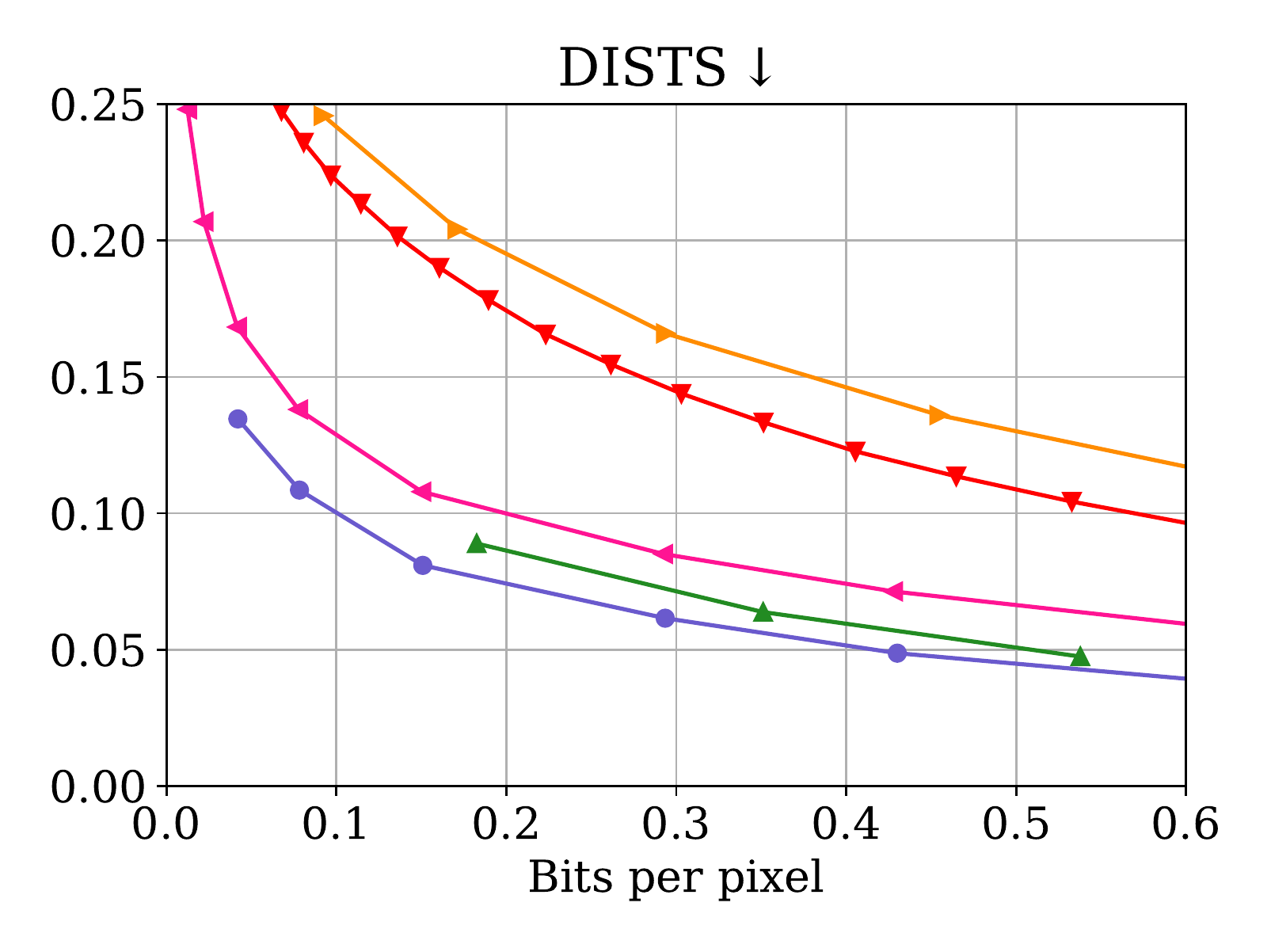}
    \end{subfigure}
    \caption{Plots of metrics for different bitrates on the Kodak dataset. In this case, we are unable to calculate FID or KID as the Kodak dataset has too few images (24) to yield useful metrics. Nonetheless, we can observe that \Ours achieves similar distortion values to \hific across the different bitrates.
    }
    \label{fig:kodak_rate_plots}
\end{figure*}

In Figure~\ref{fig:kodak_rate_plots} we show results on the Kodak dataset.
For this case, there are too few patches to calculate distributional metrics like FID and KID.
Nonetheless, the distortion metrics corroborate the performance of \Ours compared other competing methods with \Ours achieving lower distortion values than \hific at equivalent bitrates.

\section{Investigation of ImageNet feature alignment}
It has recently been demonstrated that there is a potential perceptual null space in FID due to the use of Inception V3 features in the calculation of the metric~\citep{kynkaanniemi2022role}.
Essentially, \citet{kynkaanniemi2022role} demonstrates that great improvements in FID scores can be gained by aligning images with ImageNet features without improving their perceptual quality.
\citet{kynkaanniemi2022role} also demonstrates that FID calculation with other feature extracting backbones (such as those from SSL models like SwAV~\citep{caron2020unsupervised} are not as susceptible to this effect.

\begin{figure}[ht]
    \begin{center}
    \centering
    
      \begin{subfigure}{0.8\textwidth}
        \includegraphics[width=\textwidth]{figures/rate_legend.eps}
    \end{subfigure}
    
    \begin{subfigure}{0.33\textwidth}
    \includegraphics[width=\textwidth]{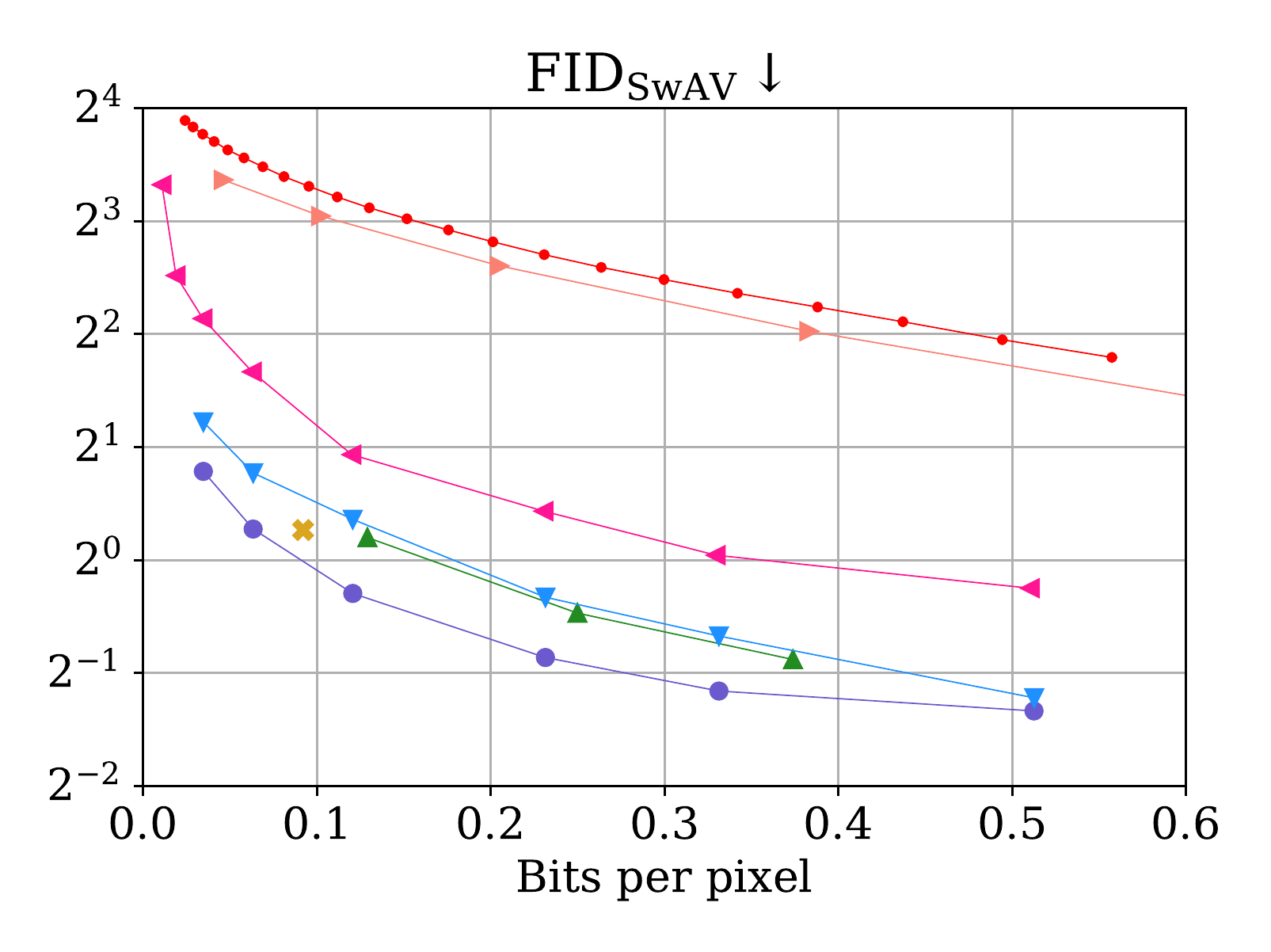}
    \subcaption{CLIC2020 test}
    \end{subfigure}
    \begin{subfigure}{0.33\textwidth}
    \includegraphics[width=\textwidth]{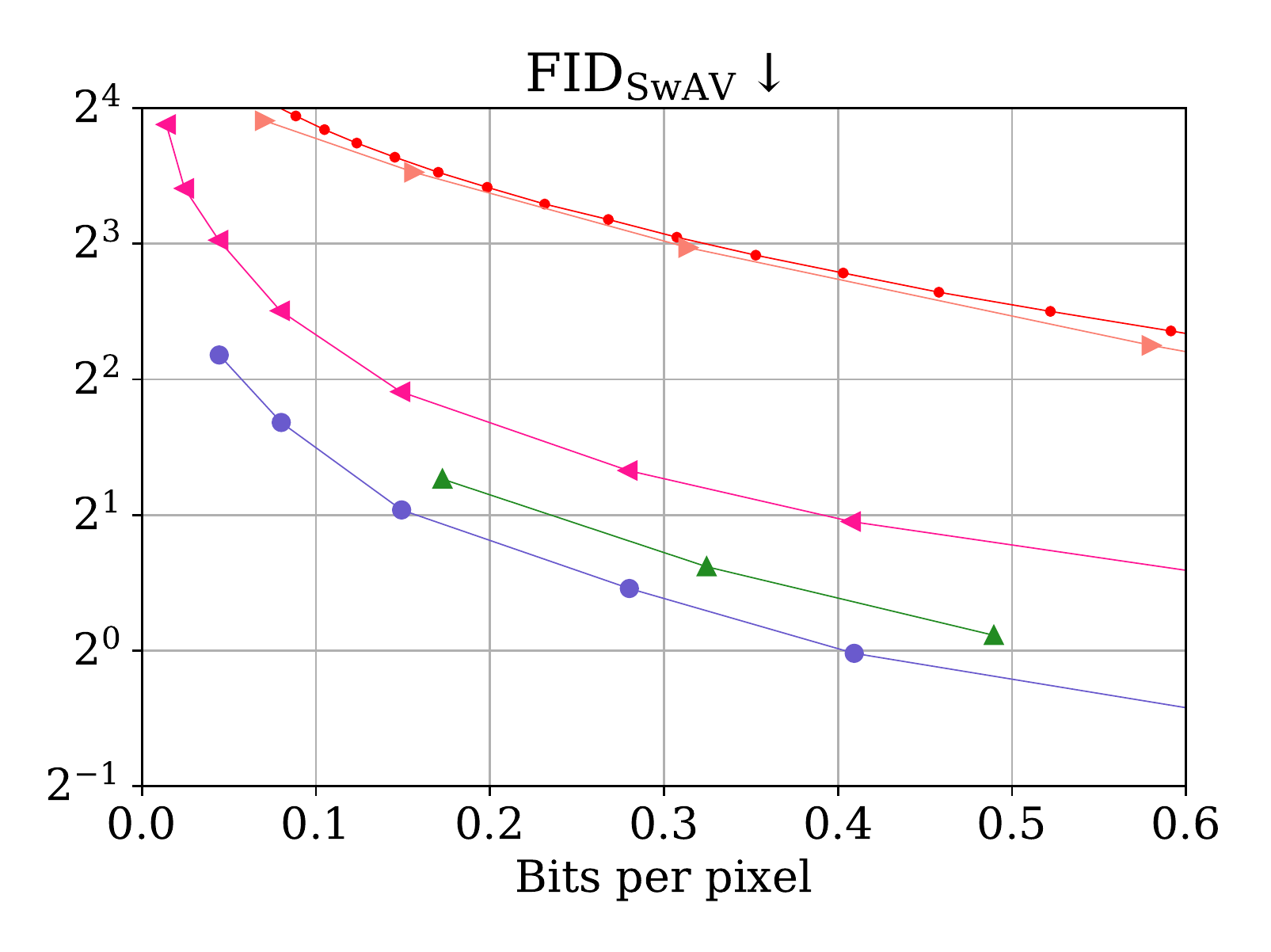}
    \subcaption{DIV2K validation}
    \end{subfigure}
    \caption{Rate-fidelity plots with claculation of FID using a ResNet50 trained via the SwAV self-supervised method~\citep{caron2020unsupervised}. Similarly to Figures~\ref{fig:rate_plots} and \ref{fig:div2k_rate_plots}, \Ours acquires higher statistical fidelity at all bitrates than the competing methods. This indicates that the improvements of \Ours arise from effects beyond ImageNet class alignment.
    }
    \label{fig:swavfid}
    \end{center}
\end{figure}

This could be a problem for our method, as we use VGG for training the labeling function $u(x)$.
This could mean that the improvement mechanism for our results might be based on ImageNet class alignment rather than improvements in image quality. For this reason, we recalculated FID scores on CLIC2020 and DIV2K using a SwAV ResNet50 background, with results in Figure~\ref{fig:swavfid}.
Figure~\ref{fig:swavfid} demonstrates our results are still upheld when using the SwAV FID extractor, indicating that our metrics improvements arise from effects beyond simple ImageNet class alignment.

\end{document}